\documentclass[12pt]{iopart}

\usepackage{graphicx}
\usepackage{xcolor}
\usepackage{cite}
\usepackage{relsize}
\usepackage{amssymb}
\usepackage{subcaption}
\usepackage{hyperref}
\usepackage{pdfrender}
\usepackage{float}

\newcommand{\NFP}{n_{\mathrm{FP}}}

\begin{document}

\title[SLAMs: Stellarator Linking Axisymmetric Mirrors]{Stellarators Linking Axisymmetric Mirrors \\
Part 1: Coil Design, MHD Equilibrium, and Physics Metrics}

\author{R.~Gaur$^1$, D.~Patch$^1$, R.~Babin$^{2,4}$, F.~Hindenlang$^2$, X.~Chu$^3$, T.~Qian$^1$\footnote{corresponding author: Tony Qian, tqian23@wisc.edu}}

\address{$^1$ Department of Physics, University of Wisconsin-Madison, Wisconsin, USA}
\address{$^2$ Max Planck Institute for Plasma Physics, Germany}
\address{$^3$ Princeton Plasma Physics Laboratory, Princeton, New Jersey, USA}
\address{$^4$ Department of Mathematics, Technical University of Munich, Germany}
\ead{rgaur@wisc.edu, rgaur@terpmail.umd.edu}
\vspace{10pt}
\begin{indented}
\item Oct 2026 
\end{indented}

\begin{abstract}
We present a computational design study of a stellarator–mirror hybrid concept, Stellarators Linking Axisymmetric Mirrors (SLAMs), in which an optimized poloidally omnigenous (OP or QI) stellarator serves as the rotational-transform-generating element that links long, axisymmetric (circular cross-section) magnetic mirrors. Starting from an OP precursor configuration obtained from the~\texttt{DESC} omnigenity database, we construct hybrid coil sets for field period $\NFP = 2$ SLAMs by separating the OP modular coils along the device midplane and inserting axisymmetric planar mirror coils into the resulting gap. Field-line tracing shows that nested vacuum flux surfaces survive this extrusion across a range of mirror-section lengths, with the cross-section qualitatively preserved at the point where the mirror is inserted. We use the field-line tracing data to map the shape of a surface and use the~\texttt{GVEC} fixed-boundary solver to calculate an ideal magnetohydrodynamic (MHD) equilibrium and evaluate physics metrics such as the neoclassical transport proxy $(\epsilon_{\rm{eff}}^{3/2})$, the vacuum magnetic well ($W$), and the gyrokinetic heat fluxes ($Q$), providing a first-principles analysis of such concepts using modern stellarator analysis tools.
\end{abstract}

\vspace{2pc}
\noindent{\it Keywords}: stellarator-mirror hybrids, stellarator optimization, stability, transport

\section{Introduction}


Magnetic-mirror and stellarator schemes sit at opposite ends of the magnetic-confinement design space. Axisymmetric mirrors have simple linear geometry, achieve high local $\beta$ in their low-field central cell, and have recently demonstrated high-field operation at 17~T using HTS magnets~\cite{Endrizzi2023_WHAM}, with the engineering path to a break-even axisymmetric mirror device set out by Forest \etal~\cite{Forest2024_BEAM}. Their two persistent physics limitations are the open loss cone, which drives parallel particle losses despite multiple-mirror plugging~\cite{BurdakovPostupaev2018_MultipleMirror}, helical-mirror flow drive~\cite{Beklemishev2013_Helicoidal, Sudnikov2017_SMOLA}, and edge-bias rotation~\cite{Bagryansky2019_LinearTraps}, and the MHD interchange instability of the unfavorably-curved central cell~\cite{RosenbluthLongmire1957_Stability}, which earlier devices controlled by minimum-B coil geometries or by providing favorable curvature beyond the mirror throat as in the gas-dynamic trap~\cite{MirnovRyutov1979_GDT, ivanov2017GasDynamicTrap}. Current high-field experiments achieve stabilization through some combination of finite-Larmor-radius effects, sheared edge rotation, and an energetic-ion population~\cite{Endrizzi2023_WHAM}. Optimized stellarators face the opposite problem, they close the loss cone by construction, can be designed for ideal-MHD stability, and admit configurations with vanishing bootstrap current and near-omnigenous drift confinement~\cite{HelanderNuhrenberg2009_QIBootstrap, goodman2023constructing, goodman2024quasi}, at the price of non-planar coils, non-axisymmetric vacuum vessels, and out-of-plane mechanical loads that drive the capital cost per unit plasma volume well above that of an equivalent mirror. A nuclear reactor that combines the engineering simplicity of the mirror with the closed field lines and drift confinement of the stellarator would therefore be of practical interest. However, whether such a device can be built without sacrificing the properties that justify modern stellarator design in the first place is an unanswered question.

This vein has been pursued in several lines of work since the 1960s, in parallel with the development of the theoretical framework for omnigenous fields. The notion of an omnigenous field, one in which the bounce-averaged radial drift of trapped particles vanishes, was formalized by Hall and McNamara~\cite{HallMcNamara1975_Omnigenous} and Catto and Hazeltine~\cite{CattoHazeltine1981_OmnigenousEq}, following earlier explicit constructions of closed isodynamic fields by Palumbo~\cite{Palumbo1968_Isodynamic}. Non-symmetric omnigenous solutions of finite extent were given by Cary and Shasharina~\cite{CaryShasharina1997_HelicalConfinement}, and the poloidally omnigenous subfamily --- omnigenous configurations with poloidally-closed contours of $|\bi{B}|$ --- was introduced by Gori \etal~\cite{Gori1996_QI}. 

In parallel, the engineering architecture for a stellarator–mirror hybrid was set out in the DRAKON proposal of Glagolev \etal~\cite{Glagolev1981_DRAKON}, two long straight mirrors joined by curvilinear elements (CRELs) such that the bulk plasma resides in low-field axisymmetric cells while the loss cone is closed by toroidal connection. Glowienka~\cite{Glowienka1981_EBT} designed a set of toroidally connected magnetic mirrors as a part of the Elmo Bumpy Torus (EBT) experiment. Moiseenko \etal~\cite{Moiseenko2014_FissionFusion} studied a stellarator–mirror hybrid for fission–fusion applications, realizing a local magnetic mirror in the Uragan-2M $l=2$ torsatron by switching off a toroidal-field coil, and concluded that the resulting mirror cell loses the symmetry favorable for trapped-particle confinement. Mazzucato~\cite{Mazzucato2020_MirrorReactor} described a racetrack reactor in which two 200~m magnetic mirrors of mirror ratio $M=2$ are connected by semicircular sections carrying $l=2$ helical windings that supply rotation and shear, projecting $\sim 13$~GW of fusion power at ITER-like density and temperature. Strickler \etal~\cite{Strickler2004_QPS} and Spong and Harris~\cite{SpongHarris2010_QPS} designed quasi-poloidal stellarator configurations that turned out to be remarkably close to a racetrack shape. Recently, Feng \etal~\cite{Feng2021_LinkedMirror} proposed a related linked-mirror device in which rotational transform is generated by a finite angle $2\Theta$ between two non-parallel straight mirror sections, a modern reinterpretation of the Figure-8 stellarator~\cite{Spitzer1958_StellaratorConcept}, and reported good single-particle and neoclassical confinement in vacuum, including for ripple-trapped orbits. The theoretical basis underlying these proposals is the observation by Shafranov \etal~\cite{Shafranov1999_QSStellaratorsMirrors} that the property responsible for good guiding-center confinement in both poloidally-closed mirrors and modern stellarators is the same quasi-symmetry, or, more generally, omnigenity, so that optimized mirrors and poloidally omnigenous stellarators can in principle be joined without disrupting drift confinement. 
We refer to the resulting class of configurations as Stellarators Linking Axisymmetric Mirrors, or SLAMs. 

This paper is organized as follows, in section~\ref{sec:cost-benefit} we present a cost-benefit analysis of a SLAM configuration and its advantage over an equivalent stellarator. In section~\ref{sec:ideal-MHD}, we introduce the fundamentals of an ideal MHD equilibrium, how it relates to physics metrics, and briefly introduce the~\texttt{DESC}~\cite{DudtKolemen2020_DESC, Panici2023_DESCSuite, conlin2023desc} and \texttt{GVEC}~\cite{Hindenlang2026_GVEC} codes. In section~\ref{sec:DESC}, we propose a novel way to create a stellarator-mirror hybrid coilset given a QI coilset, demonstrate the existence of flux surfaces in the hybrid, and perform QFM (Quadratic Flux Minimizing) surface optimization to axisymmetrize the mirror in the pre-SLAM coilset. Next, we develop the $\texttt{FTGen}$ algorithm to map the flux surfaces in multiple coordinate systems.
Since~\texttt{DESC} is limited to using cylindrical coordinates, which are ill-suited for the SLAM geometry, we use~\texttt{GVEC} with curve-following coordinates to solve the fixed-boundary problem using the surface data obtained from the previous section. Using the \texttt{GVEC} equilibrium, we calculate the effective ripple, magnetic well, and gyrokinetic heat flux and compare these properties for the SLAM, pre-SLAM, and the original QI design. Finally, in section~\ref{sec:conclusions}, we conclude our work and discuss the directions in which it will be extended.


\section{Cost-benefit analysis of pursuing the SLAM}
\label{sec:cost-benefit}
The engineering case for pursuing SLAM rests on a volumetric scaling argument that we set out here as motivation rather than as a quantitative cost model. Consider two devices of equal total plasma volume: a conventional poloidally omnigenous (OP), also referred to as a Quasi-isodynamic (QI) stellarator of volume $V_s^0$, and a SLAM device consisting of a stellarator linker of volume $V_s$ joined to an axisymmetric mirror cell of volume $V_m$, with $V_s + V_m = V_s^0$. Treating the capital cost of each section as proportional to its enclosed plasma volume, a zeroth-order approximation justified by the observation that the dominant cost drivers (magnets, vacuum vessel, support structure, blanket) all scale roughly with surface area or enclosed volume, we write the total costs as $C_s^0 = k_s V_s^0$ for the conventional stellarator and $C_\mathrm{SLAM} = k_s' V_s + k_m V_m$ for the hybrid, where $k_s$ is the cost per unit plasma volume of a free-standing stellarator section, $k_m$ is the corresponding quantity for an axisymmetric mirror section, and $k_s'$ is the cost per unit volume of the stellarator section \emph{within} a SLAM device, which may differ from $k_s$ if, for example, neutron-flux localization in the mirror cell relaxes the shielding requirement and permits modular coils to be brought closer to the plasma. Three dimensionless ratios then capture the essential dependence
\begin{equation}
    \hat{\alpha} \equiv \frac{k_s}{k_m} \geq 1, \qquad
    \hat{\gamma} \equiv \frac{k_s'}{k_s} \leq 1, \qquad
    V_r \equiv \frac{V_m}{V_s},
\end{equation}
where $\hat{\alpha}$ measures the cost-per-volume penalty of three-dimensional stellarator engineering relative to axisymmetric mirror engineering, $\hat{\gamma}$ is the cost-reduction factor available to the stellarator section once it is embedded in a SLAM, and $V_r$ is a design choice. In these variables, the ratio of conventional to SLAM cost at equal plasma volume reduces to
\begin{equation}
    C_r \equiv \frac{C_s^0}{C_\mathrm{SLAM}} = \frac{1 + V_r}{\hat{\gamma} + V_r/\hat{\alpha}} > 1.
\end{equation}
The structure of $C_r$ is transparent in two limits, first a limit where $V_r \to 0$ gives $C_r \to 1/\hat{\gamma}$, the saving from neutron-flux relaxation alone, while $V_r \to \infty$ gives $C_r \to \hat{\alpha}$, the full per-volume saving. Cost advantage $C_r > 1$ therefore requires either $\hat{\alpha} > 1$ (mirror sections are intrinsically cheaper per cubic meter of plasma) or $\hat{\gamma} < 1$ (the stellarator section benefits from neutron-flux localization), and, provided $\hat{\alpha}\hat{\gamma} > 1$, is amplified by increasing $V_r$ until the asymptotic limit $\hat{\alpha}$ is approached. 

Two further relations link this scaling to plasma physics. Conservation of toroidal flux along a flux tube, $B A_\perp = \mathrm{const}$, gives $V_r = M_r L_r$, where $M_r \equiv B_s/B_m$ is the mirror ratio between the stellarator and mirror sections and $L_r$ is the ratio of mirror-to-stellarator arc lengths, and the local plasma $\beta = 2\mu_0 p / B^2$ at fixed flux-surface pressure $p$ scales as $\beta_m = M_r^2 \beta_s$, so that operating at large mirror ratio simultaneously raises plasma volume and magnetic-utilization efficiency. We emphasize that the equations above are scaling estimates, not engineering cost models---values of $\hat{\alpha}$, $\hat{\gamma}$, $V_r$ and $M_r$ in the favorable regime would all need to be substantiated by a full engineering analysis---but they identify the parameter range in which a stellarator--mirror hybrid is worth pursuing and they expose three operating tiers, a baseline tier with $V_r > 0$ at the natural OP mirror minimum and $M_r$ of order unity (SLAM-A), a higher-magnetic-utilization tier with $M_r \gg 1$ and isotropic pressure (SLAM-B), and a tier with $M_r \gg 1$ and anisotropic pressure $p_\perp \neq p_\parallel$ that localizes fusion power density to the mirror cell (SLAM-C). The present paper addresses SLAM-A as the necessary first step.

\section{Ideal MHD equilibrium}
\label{sec:ideal-MHD}
In this section, we briefly explain how we define and calculate an ideal MHD equilibrium in a closed, toroidal device.

A divergence-free magnetic field $\bi{B}$ can be written in the Clebsch form~\cite{d2012flux}
\begin{equation}
    \bi{B} =   \bi{\nabla} \psi \times \bi{\nabla}\alpha.
    \label{eqn:Div-free-B2}
\end{equation}
We will focus on solutions whose magnetic field lines lie on closed nested toroidal surfaces, known as flux surfaces. We label these surfaces using the enclosed toroidal flux $\psi$ or the normalized radius $\rho = \sqrt{\psi/\psi_{\rm{b}}}$, where $\psi_{\rm{b}}$ is the toroidal flux enclosed by the boundary. On each flux surface, the lines of constant $\alpha$ coincide with the magnetic field lines. Thus, $\alpha$ is known as the field line label. We define $\alpha = \theta_{\rm{PEST}} - \iota (\zeta-\zeta_0)$, where $\theta_{\rm{PEST}}$ is the $\mathrm{PEST}$~\cite{grimm1983ideal} straight-field-line poloidal angle, $\zeta$ is the generalized cylindrical toroidal angle, $\zeta_0$ is a constant, and 
\begin{equation}
    \iota = \frac{\bi{B}\cdot \bi{\nabla}\theta_{\mathrm{PEST}}}{\bi{B}\cdot \bi{\nabla}\zeta},
\end{equation}
is the pitch of the magnetic field lines on a flux surface, known as the rotational transform. Using the Clebsch form of the magnetic field, we solve the ideal MHD force balance equation
\begin{equation}
    \bi{j} \times \bi{B} = \bi{\nabla} p,
    \label{eqn:ideal-MHD-force-balance}
\end{equation}
where the plasma current $\bi{j} = (\bi{\nabla} \times \bi{B})/\mu_0$ from Amp\`ere's law, $p$ is the plasma pressure and $\mu_0$ is the vacuum magnetic permeability. Unlike a toroidally-symmetric case, for stellarators, we have to solve~\eref{eqn:ideal-MHD-force-balance} as an optimization problem. We can achieve this with either the $\texttt{DESC}$ or $\texttt{GVEC}$ 3D ideal MHD equilibrium solvers. These codes can solve for an equilibrium and compute multiple objectives such as ideal MHD stability proxies, quasisymmetry, and many more. The details of the~\texttt{DESC} package, which can also be used to optimize the equilibrium for various objectives, are provided in these references~\cite{dudt2020desc, panici2023desc, conlin2023desc, dudt2023desc} and of the~\texttt{GVEC} code in Hindenlang \etal~\cite{Hindenlang2025_GFrame,Hindenlang2026_GVEC}.

In this paper, we will use~\texttt{DESC} to design SLAM-A coils and~\texttt{GVEC} to solve a fixed-boundary equilibrium resulting from the SLAM-A coils. \texttt{GVEC} assumes nested flux surfaces and takes the boundary, the total toroidal flux, the pressure profile, and either the rotational transform or the toroidal-current profile as inputs. The flux surfaces are defined locally with respect to a three-dimensional, closed guiding curve instead of being defined with respect to a global coordinate system. The numerical optimization process is similar to the Variational Moments Equilibrium Code (\texttt{VMEC})~\cite{HirshmanWhitson1983_VMEC} and minimizes the total MHD energy. The output is an equilibrium with nested flux surface shapes from which one can calculate various physics metrics. Note that, in this work we will use the cylindrical coordinate system $(R, \phi, Z)$ to represent surfaces and set the toroidal angle $\zeta = \phi$, unless stated otherwise.

In the next section, we will use~\texttt{DESC} and the precursor OP equilibrium and coilset, design the resulting stellarator-mirror hybrid coilset, and trace field lines to parametrize flux surfaces from the SLAM-A coilset. 

\section{Creating a SLAM-A coilset with DESC}
\label{sec:DESC}
In this section, we will describe in detail how we used~\texttt{DESC} to design coils for a SLAM-A equilibrium. Unlike the conventional order of first designing an equilibrium (stage-one optimization) and then coils (stage-two optimization), we first design a stellarator-mirror hybrid coilset from a poloidally omnigenous (OP) coilset taken from the~\texttt{DESC} omnigenity database~\cite{Gaur-DPP23}. We spatially dissect the OP coilset and add planar mirror coils between the stellarator coils to form a stellarator-mirror hybrid. We then trace the field lines resulting from the hybrid coilset to demonstrate the existence of nested flux surfaces. Since the hybrid coilset does not lead to a circular mirror region, we use~\texttt{DESC}'s Quadratic Flux Minimizing (QFM) surface optimization to modify a subset of the mirror coils to ensure that the mirror is axisymmetric (circular) without affecting the stellarator surface shape. Finally, we trace a single field line to map a flux surface created by the coils (in different coordinate systems) to use as an input to the equilibrium code~\texttt{GVEC}.

\subsection{Creating the stellarator mirror hybrid coilset}
\begin{figure}[h]
    \centering
    \begin{subfigure}[b]{0.27\textwidth}
    \centering
        \includegraphics[width=\textwidth]{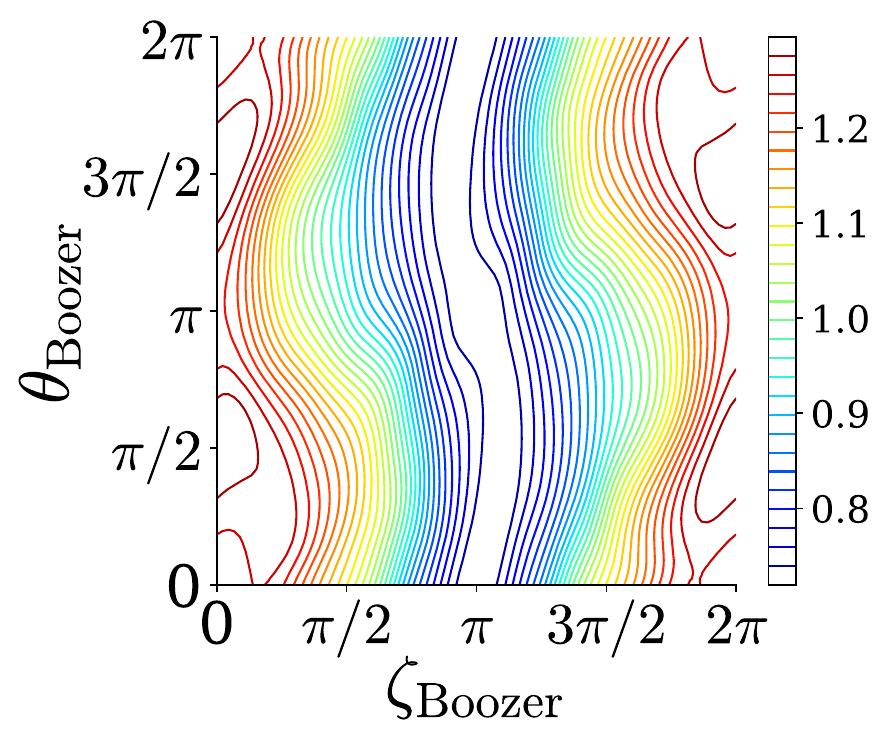}
        \caption{$|\bi{B}|$($\rho=1.0$)}
        \label{subfig:Boozer}
    \end{subfigure}
    \begin{subfigure}[b]{0.32\textwidth}
        \centering
        \includegraphics[width=\textwidth, trim={0mm 0mm 15mm 45mm}, clip]{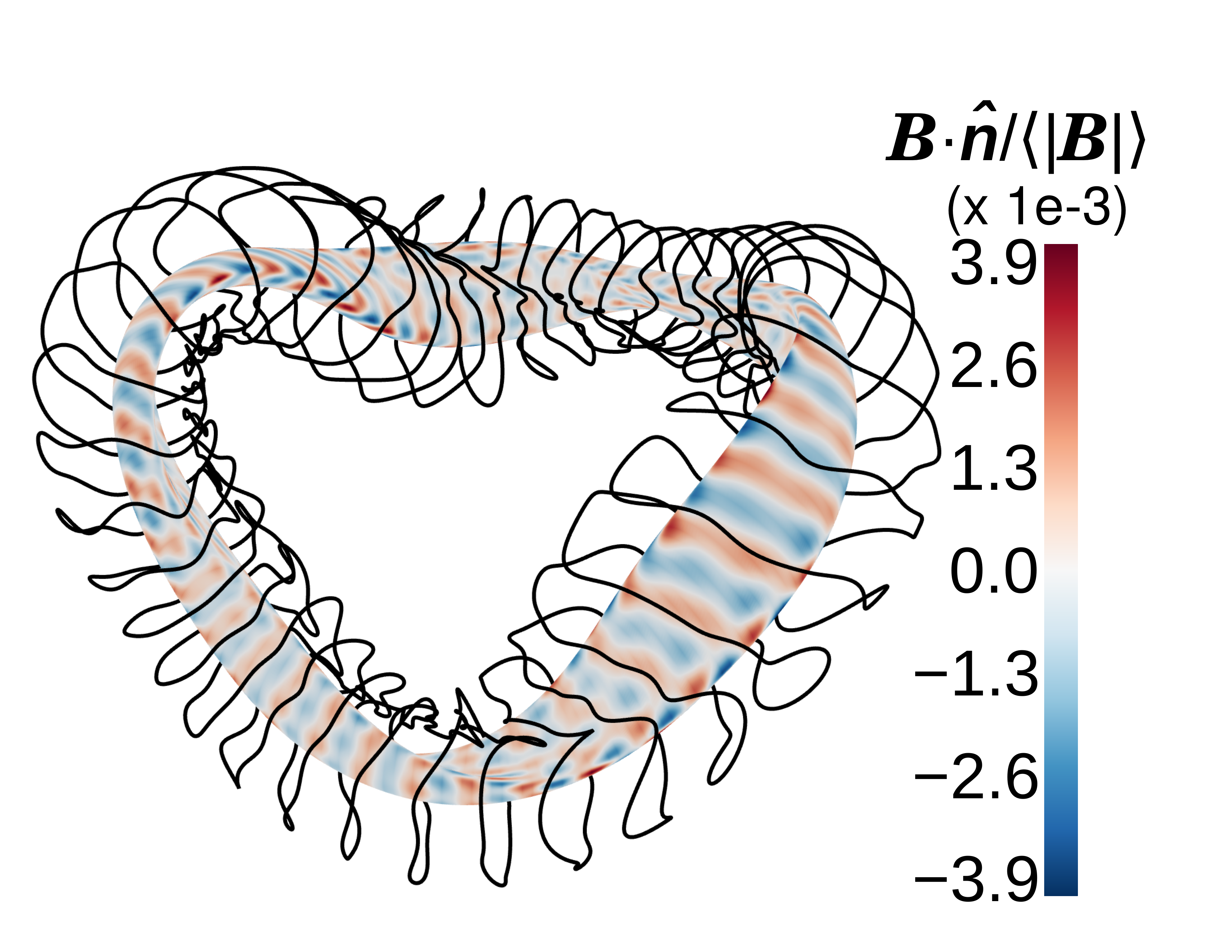}
        \caption{$\bi{B}\!\cdot\!\hat{\bi{n}}/\langle | \bi{B} | \rangle$ and coils}
        \label{subfig:Bdotn_coils}
    \end{subfigure}
    \begin{subfigure}[b]{0.37\textwidth}
        \centering
        \includegraphics[width=\textwidth, trim={0mm 0mm 2mm 0}, clip]{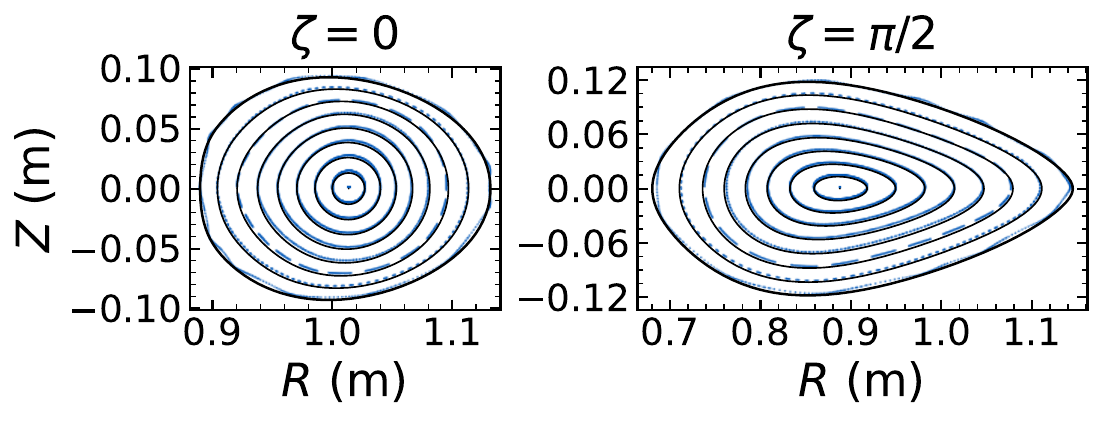}\\[5mm]
        \caption{Poincar\'e trace}
        \label{subfig:Poincare}
    \end{subfigure}
    \caption{A $\NFP = 2$ vacuum OP equilibrium and its corresponding coilset from the~\texttt{DESC} omnigenity database~\cite{2023APS_DESC_omnigenity_database}. Figure~\ref{subfig:Boozer} represents the field strength in Boozer coordinates. Figure~\ref{subfig:Bdotn_coils} plots the OP coilset and the normalized field error $\bi{B}\cdot\hat{\bi{n}}/\langle | \bi{B} | \rangle$ on the equilibrium boundary, where $\langle \rangle$ is a surface average. Figure~\ref{subfig:Poincare} traces the field lines from the coils and compares them with equilibrium flux surfaces. The coilset field-line trace matches well with the flux surfaces.}
\label{fig:QI-equilibrium}
\end{figure}

We take an optimized two-field-period OP equilibrium and its coilset from the~\texttt{DESC} omnigenity database. The field period $\NFP$ is the number of discrete identical sections that make up the stellarator. The equilibrium is generated using stage-one optimization where the boundary shape is varied to ensure favorable physics properties and the coilset is designed as a stage-two optimization, where the boundary is fixed and coils are generated to minimize the normal field error $\bi{B}\cdot\hat{\bi{n}}$ (where $\hat{\bi{n}} = \bi{\nabla} \rho/| \bi{\nabla} \rho|$) on the boundary. The coilset and properties of the equilibrium are presented in figure~\ref{fig:QI-equilibrium}.

\begin{figure}[h]
    \centering
    \begin{subfigure}[b]{0.30\textwidth}
    \centering
        \includegraphics[width=\textwidth, trim={5mm 0mm 5mm 5mm}, clip]{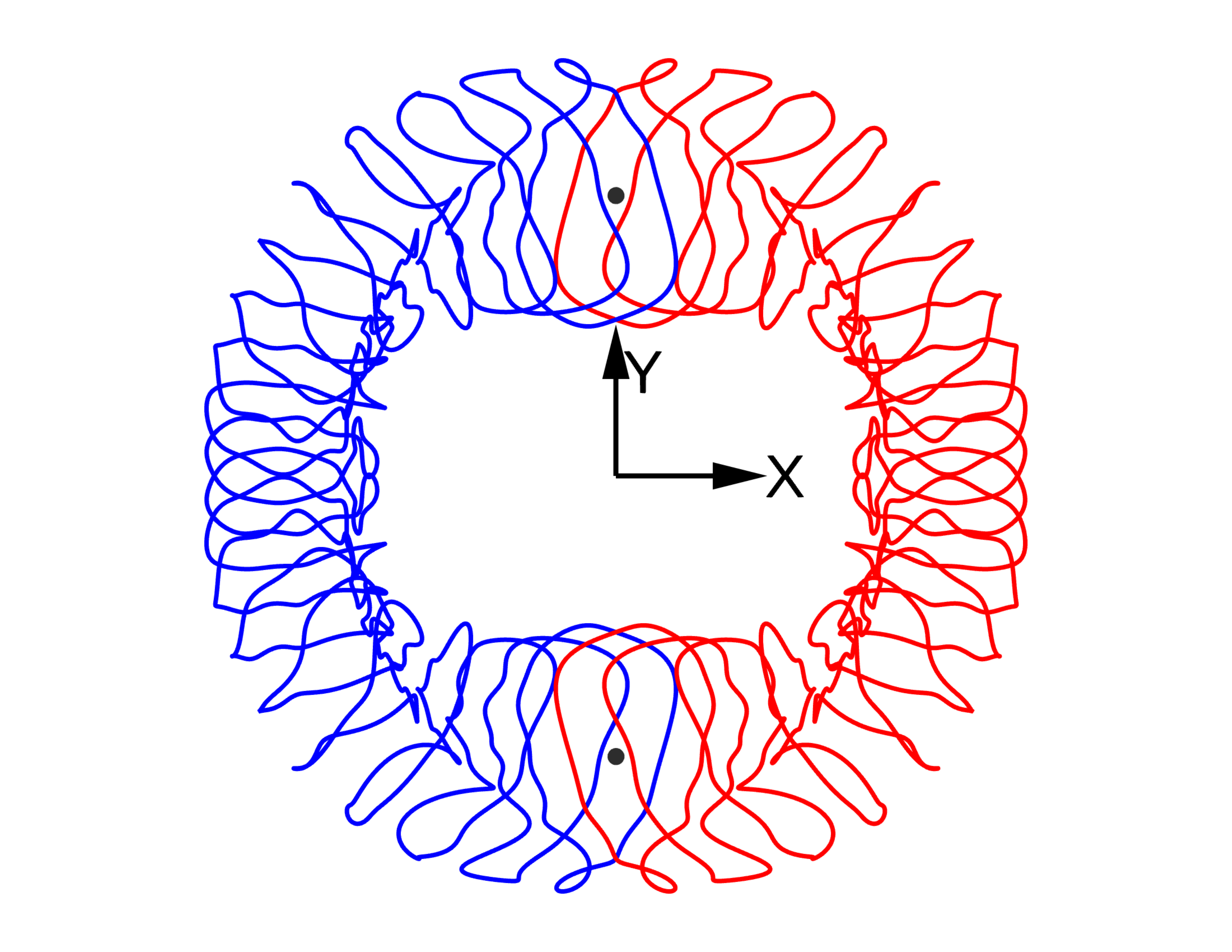}
        \caption{Original}
        \label{subfig:original-coilset}
    \end{subfigure}
    \begin{subfigure}[b]{0.315\textwidth}
        \centering
        \includegraphics[width=\textwidth, trim={5mm 0mm 5mm 5mm}, clip]{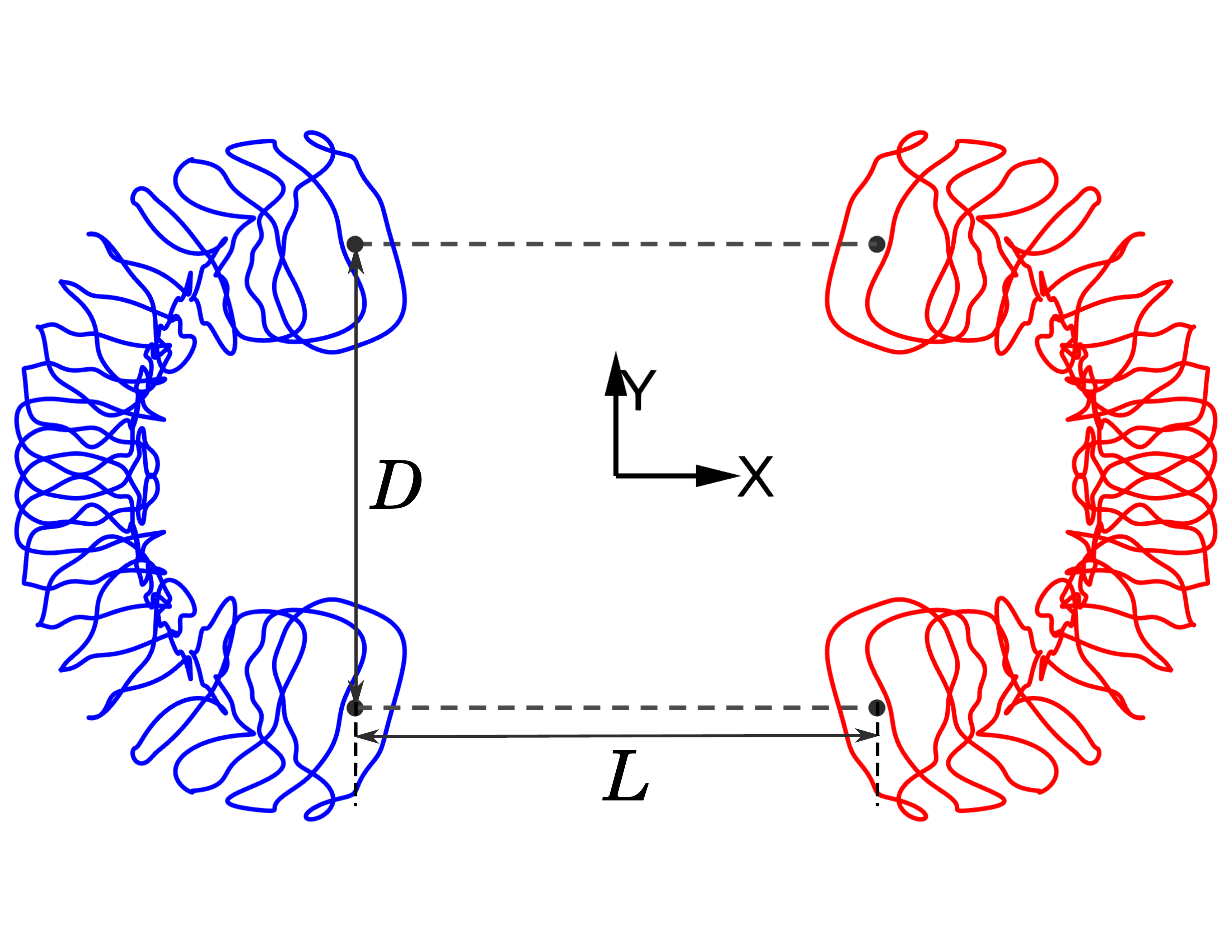}
        \caption{Separated along $X$ axis}
        \label{subfig:coilset-separation}
    \end{subfigure}
    \begin{subfigure}[b]{0.32\textwidth}
        \centering
        \includegraphics[width=\textwidth, trim={5mm 30mm 5mm 0}, clip]{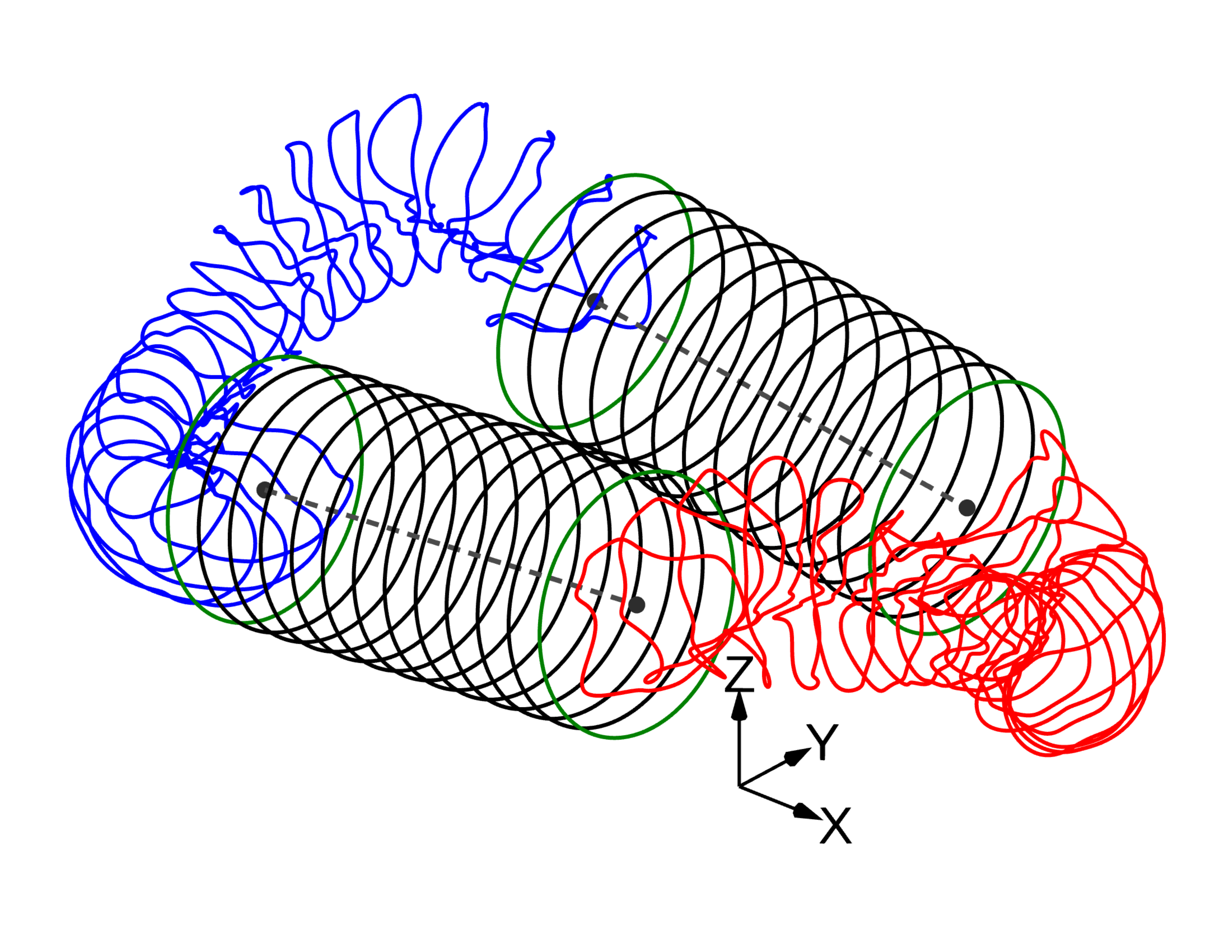}
        \caption{Stellarator-Mirror Hybrid}
        \label{subfig:SLNAM-coilset-subfig}
    \end{subfigure}
    \caption{Process of making a stellarator-mirror hybrid coilset. We start with the original OP coilset in figure~\ref{subfig:original-coilset}, separate it at the point of lowest magnetic field strength and vanishing axis curvature, along the $X$ axis by a distance $L$ in figure~\ref{subfig:coilset-separation}, and using the (dashed) line joining the magnetic axis in the original coilset as the center, place circular mirror coils in the gap as shown in figure~\ref{subfig:SLNAM-coilset-subfig}.}
\label{fig:SLNAM-coilset}
\end{figure}

To create a stellarator-mirror hybrid, we take the coils in the OP coilset, and separate them about the low-field region\footnote{One can separate the coils and add a mirror in a low-field and/or high-field region but adding a mirror in the low-field region is more advantageous from the analysis in section~\ref{sec:cost-benefit}.}, where the magnetic axis curvature vanishes. We then shift the coils and fill the gap with circular mirror coils, with all mirror coils having the same current, which is $1.5$\footnote{The mirror current in the hybrid coilset was determined based on the ratio of the effective radius of the low-field region stellarator coils and the mirror coils. The analysis in this paper works over a broad range of mirror coil currents and the choice does not affect the outcomes significantly.} times the current carried by the closest low-field stellarator coil. The process of converting a stellarator coilset to a stellarator-mirror hybrid coilset is demonstrated in figure~\ref{fig:SLNAM-coilset}.

\begin{figure}[h]
    \vspace*{-2mm}
    \centering
    \begin{subfigure}[b]{0.45\textwidth}
    \centering
        \includegraphics[
        width=0.75\textwidth, trim={2mm 8mm 8mm 50mm}, clip]{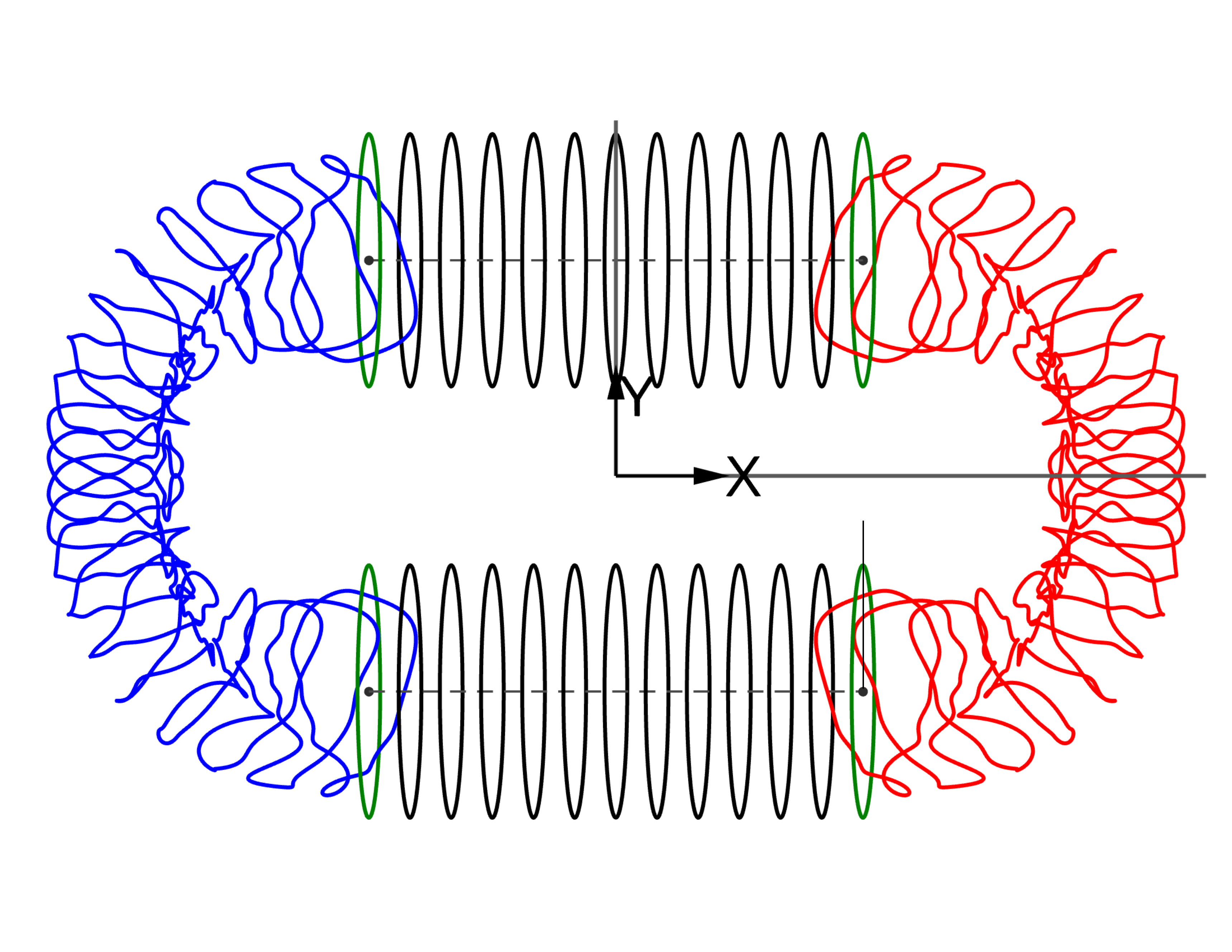}
    \end{subfigure}
    \begin{subfigure}[b]{0.45\textwidth}
        \centering
        \raisebox{-5.0mm}{
        \includegraphics[width=\textwidth, trim={2mm 8mm 8mm 50mm}, clip]{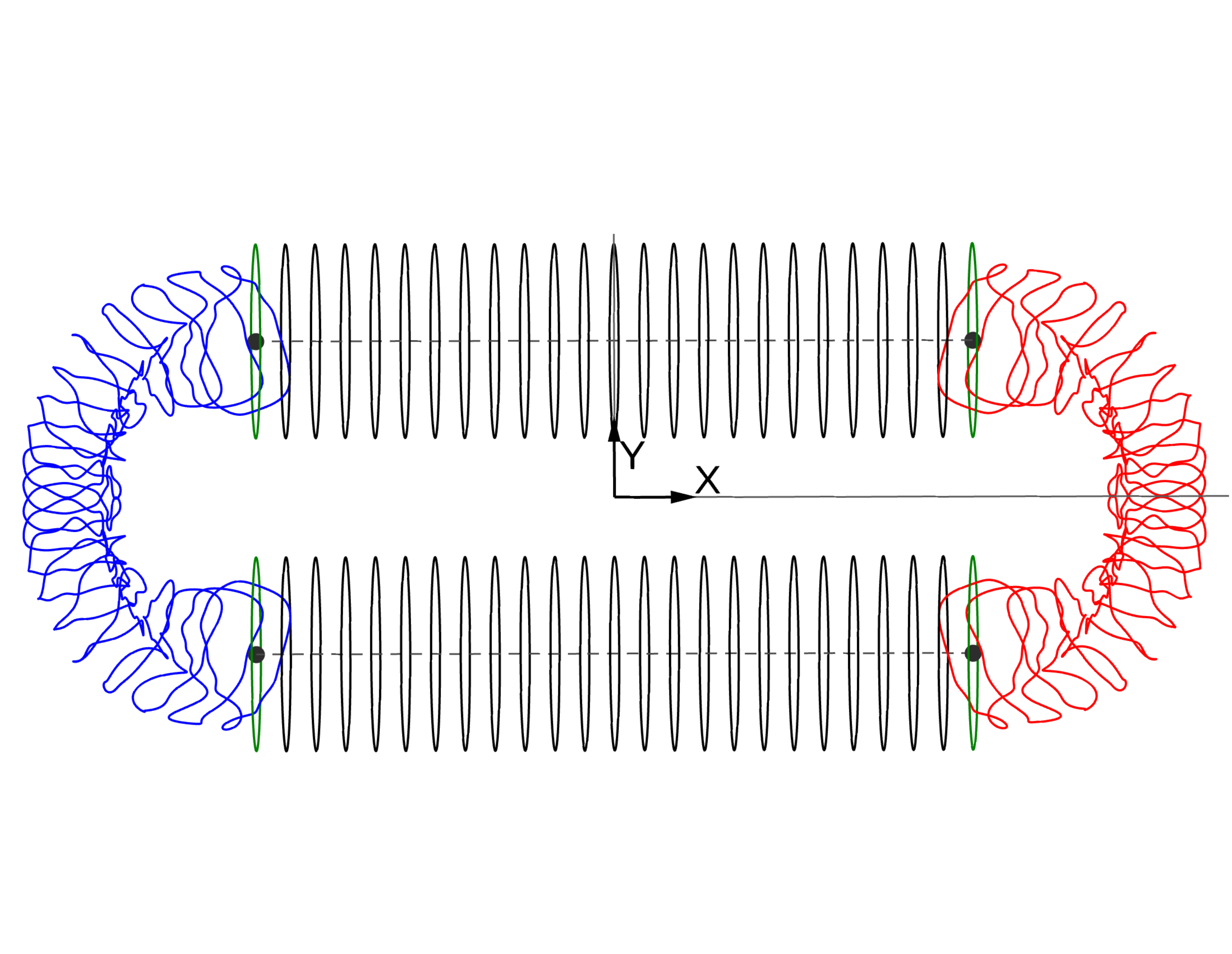}}
    \end{subfigure}\\[-5mm]
    
    \begin{subfigure}[b]{0.45\textwidth}
    \centering
        \includegraphics[width=\textwidth, trim={2mm 2mm 2mm 2mm}, clip]{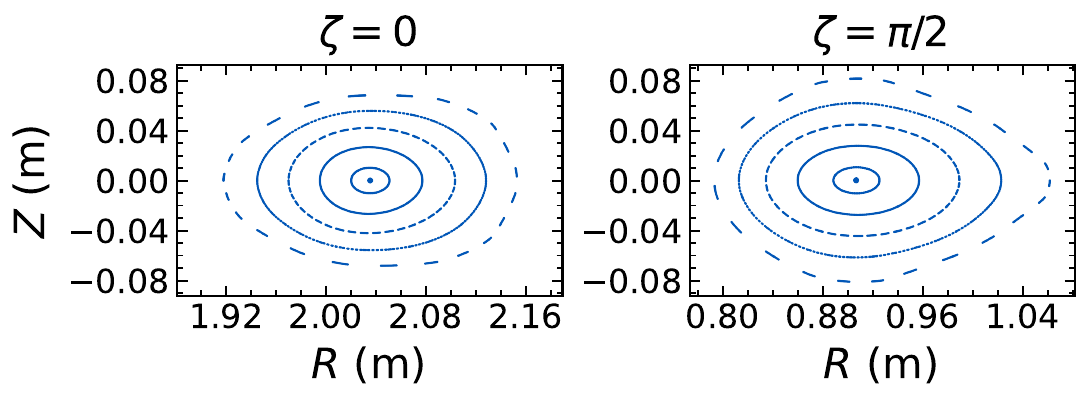}
        \caption{$L/D=1.0$}
        \label{subfig:L_over_D_1}
    \end{subfigure}
    \begin{subfigure}[b]{0.45\textwidth}
        \centering
        \includegraphics[width=\textwidth, trim={2mm 2mm 2mm 2mm}, clip]{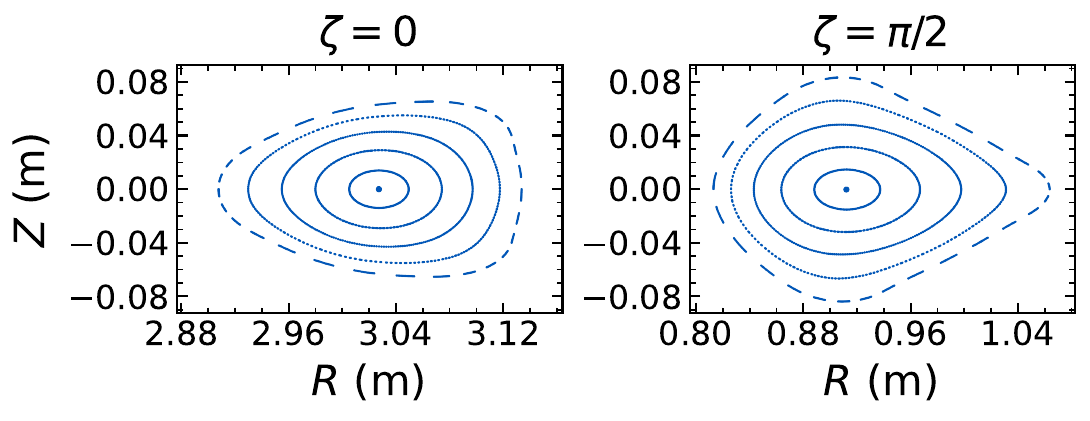}
        \caption{$L/D=2.0$}
        \label{subfig:L_over_D_2}
    \end{subfigure}
    \caption{Stellarator-mirror hybrid coilset using the OP coilset presented in figure~\ref{subfig:Bdotn_coils} with two different mirror lengths and their corresponding Poincar\'e plots on the planes where the black lines (top row) intersect the coils. Nested surfaces exist even after separating the OP coilset and adding mirror coils. Note that the cross-section in the mirror is not a circle (axisymmetric).}
\label{fig:SLNAM-Poincare}
\end{figure}

We characterize the racetrack-shaped stellarator-mirror hybrid using the ratio of the length of the mirror $L$ to the major diameter of a stellarator $D = 2 R$, $L/D$, where $R$ is the major radius of the original OP stellarator. This parameter is related to $L_r$ defined in section~\ref{sec:cost-benefit} such that $L/D = \pi L_r$. We now discuss two potential issues with the coil arrangement. First, after adding the mirror coils, there is no guarantee that the magnetic field will form nested flux surfaces.  Second, the line joining the magnetic axis from the original coilset with the mirror coil center is non-smooth and can potentially change the physics properties of the stellarator-mirror hybrid by introducing ripples in the magnetic field. We will revisit this point when we analyze the physics properties in section~\ref{sec:GVEC}. To test the first point, we trace field lines from the stellarator-mirror hybrid coilset in figure~\ref{fig:SLNAM-Poincare}.

We find that nested flux surfaces exist even after adding the circular mirror coils in the low-field region at least up to $L/D = 2$. In terms of the convention used by Feng \etal~\cite{Feng2021_LinkedMirror}, the angle between the two straight sections $2\Theta \neq 0$ for either case in figure~\ref{fig:SLNAM-Poincare}. However, since the OP coilset stays the same for both the mirror lengths, $2\Theta (L/D = 2) < 2\Theta (L/D = 1)$, and the two mirrors become more parallel as we increase the mirror length, \textit{i.e.}, $\lim_{L/D \rightarrow \infty} 2 \Theta = 0$. 

We also find that the nestedness of the stellarator-mirror hybrid is robust to the current in the mirror coils. Varying the current from $0.8$--$2.0$ times the current in the adjacent low-field coil sustains nestedness, but changes the flux surface shapes in both the stellarator and mirror regions. Interestingly, the low-field Poincar\'e cross-section strongly resembles the low-field Poincar\'e trace in the original OP coilset even after creating the hybrid. For further analysis in this paper, we choose the stellarator-mirror hybrid with $L/D = 1$.

In the next section, we will use the existence of flux surfaces in a stellarator-mirror hybrid coilset to optimize the mirror section to become circular (axisymmetric).

\subsection{Axisymmetrizing the mirror region without changing the stellarator region}
We take advantage of flux surface nestedness in the stellarator-mirror hybrid by calculating a surface on which the quadratic flux $(\bi{B}\cdot \hat{\bi{n}})^2$ (normal magnetic field) is minimized. Such a surface is also known as a Quadratic Flux Minimizing (QFM) surface. This idea was first used by Dewar \etal~\cite{DewarQFM1994} to minimize the squared action gradient, so QFM surfaces are families of periodic pseudo-orbits (of field lines) with rational iota. As the quadratic flux on a QFM surface approaches zero, it becomes identical to a flux surface\footnote{A QFM surface method can fail near an island surface, which is why it is important to check the existence of closed, island-free flux surfaces before doing QFM optimization.}.

\begin{figure}[h]
    \centering
    \begin{subfigure}[b]{0.31\textwidth}
    \centering
        \includegraphics[width=\textwidth, trim={5mm 0mm 5mm 5mm}, clip]{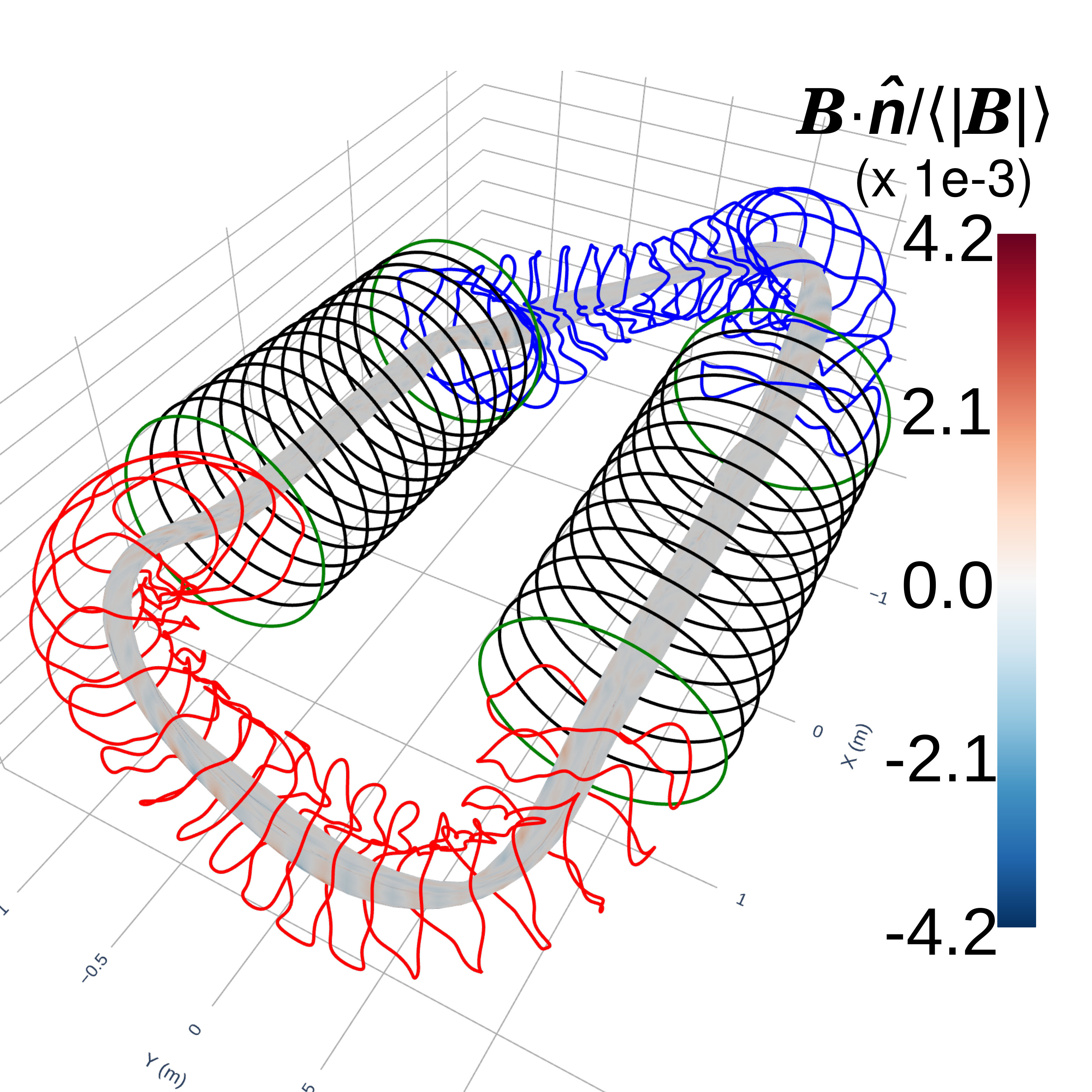}
        \caption{pre-SLAM QFM $\mathcal{S}_1$}
        \label{subfig:SLNAM-QFM}
    \end{subfigure}
    \,
    \begin{subfigure}[b]{0.31\textwidth}
        \centering
        \includegraphics[width=\textwidth, trim={5mm 0mm 5mm 5mm}, clip]{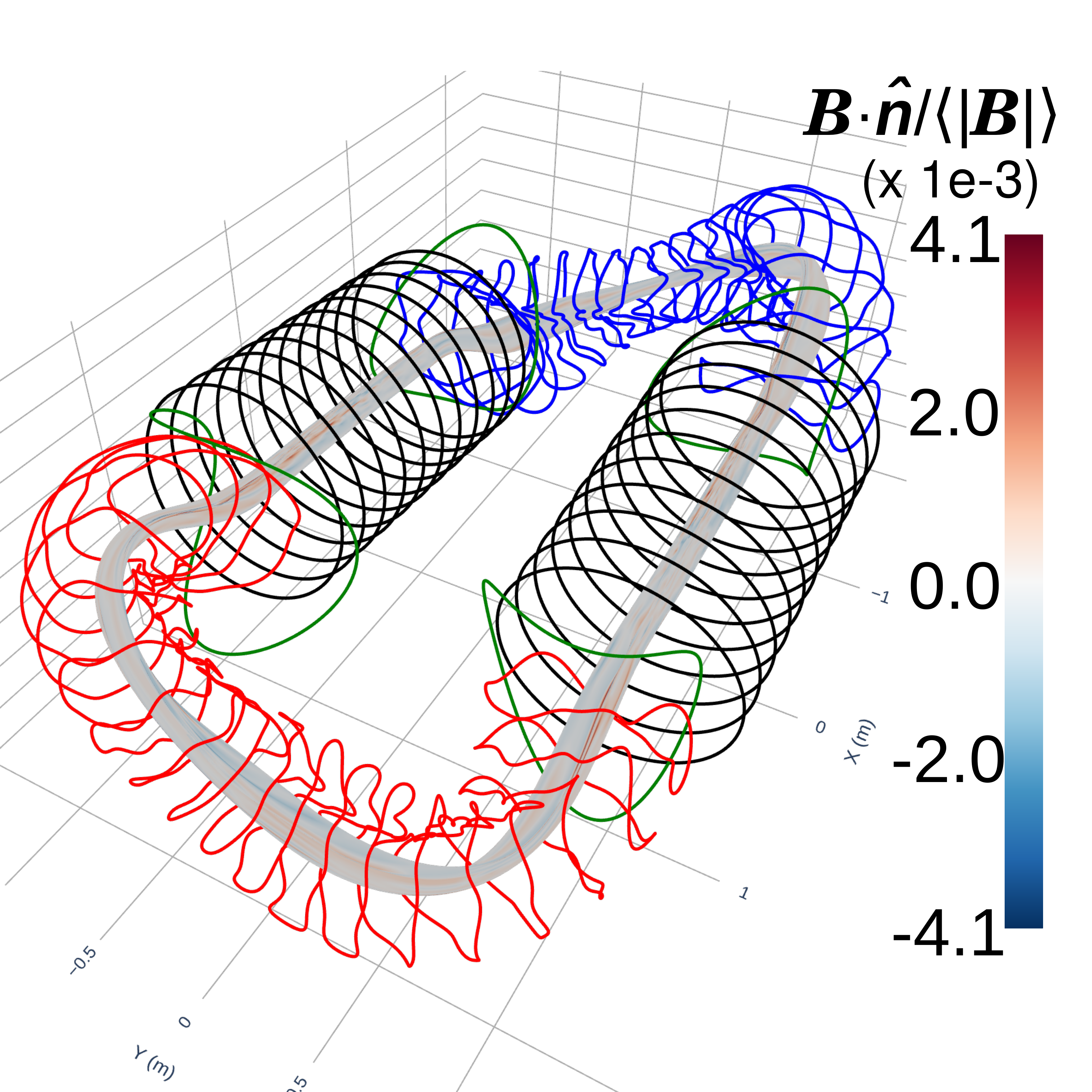}
        \caption{SLAM QFM $\mathcal{S}_2$}
        \label{subfig:SLAM-QFM}
    \end{subfigure}
    \, 
    \begin{subfigure}[b]{0.28\textwidth}
        \centering
        \includegraphics[width=\textwidth, trim={2mm 0mm 0mm 2mm}, clip]{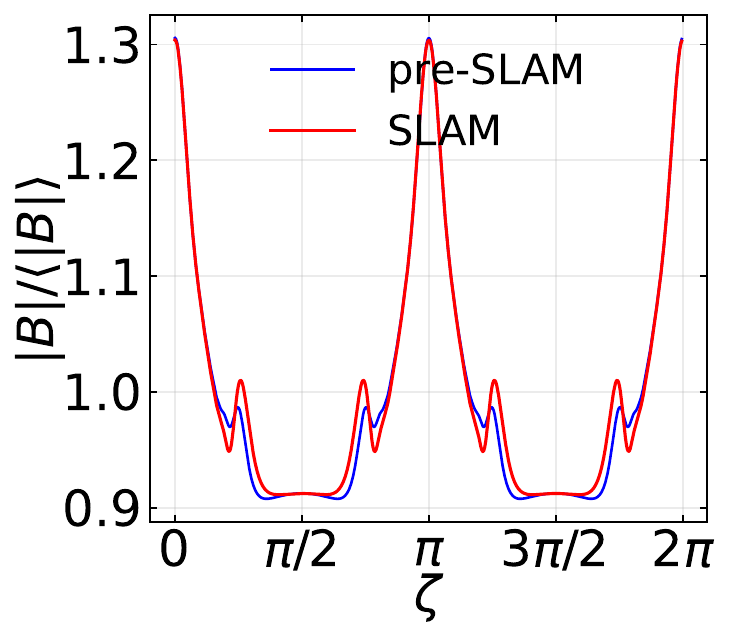}\\[3mm]
        \captionsetup{width=1.4\linewidth,margin={-0.08\linewidth,0pt}}
        \caption{Normalized $|\bi{B}|$ on-axis}
        \label{subfig:modB-on-axis}
    \end{subfigure}
    \caption{The pre-SLAM and SLAM coilsets and QFM surfaces along with the magnetic field on axis for the two coilsets. Note that only a single unique converter coil is changed which introduces a magnetic field ripple in the converter region around $\zeta = \pi/4, 3\pi/4, 5\pi/4, 7\pi/4$ as shown in figure~\ref{subfig:modB-on-axis}, while axisymmetrizing the mirror and keeping the stellarator boundary almost exactly the same. As a consequence, the $|\bi{B}|$ in the stellarator region is identical for both the pre-SLAM and SLAM coilsets.}
\label{fig:SLNAM-to-SLAM}
\end{figure}

We will first calculate the QFM surface from the initial stellarator-mirror hybrid coilset shown in figure~\ref{subfig:L_over_D_1}, parametrized using a double Fourier series with the corresponding coefficients $R_{\mathrm{b}, mn}, Z_{\mathrm{b}, mn}$ where $m$ and $n$ are the poloidal and toroidal mode numbers respectively. We allow these coefficients to change such that they minimize the objective
\begin{eqnarray}
  \mathcal{F}_{\rm{QFM}} =  w_{Q} f_{\mathrm{QF}}^2 + w_{T} f_{\psi_{\rm{b}}}^2,
\end{eqnarray}
where $f_{\mathrm{QF}}$ and $f_{\psi_{\rm{b}}}$ are objectives penalizing the quadratic flux and enclosed toroidal flux, respectively. We solve the optimization problem $\min (\mathcal{F}_{\mathrm{QFM}})$ subject to a constraint imposing a fixed toroidal flux, and find the QFM surface $\mathcal{S}_1(R_{\mathrm{b}, mn}, Z_{\mathrm{b}, mn})$. We will refer to the initial hybrid coilset along with the surface $\mathcal{S}_1$ as the pre-SLAM configuration. After solving for the pre-SLAM configuration, we re-optimize it to axisymmetrize the mirror by allowing the green coil to change its shape and create a SLAM (axisymmetric mirror region) while keeping the stellarator part of the SLAM unchanged. The objective function for the re-optimization problem is 
\begin{eqnarray}
\fl \mathcal{F}_{\rm{QFM}}^{\rm{SLAM}} = w_{Q} f_{\mathrm{QF}}^2 + w_{T} f_{\psi_{\rm{b}}}^2 + w_{D} f_{\rm{coil-coil\, dist.}}^2 + w_{L} f_{\rm{coil-length}}^2 \nonumber \\
                                 + w_{S} f_{\rm{surf-match}}^2 + w_{E} f_{\rm{SLAM-elong}}^2 + w_C f_{\rm{SLAM-center}}^2,
\end{eqnarray}
where the new objective functions penalize engineering metrics such as coil-coil separation and coil length while ensuring that the SLAM QFM surface $\mathcal{S}_2$ matches the pre-SLAM surface $\mathcal{S}_1$ in the stellarator region, the SLAM is circular in the mirror region ($f_{\mathrm{SLAM-elong}}$), and the center of the axisymmetric mirror is a straight line ($f_{\mathrm{SLAM-center}}$). These objectives are defined and motivated in detail in~\ref{app:objectives}. To create a SLAM, we allow one mirror coil per half field period (green coil in figure~\ref{subfig:L_over_D_1}), which we call a converter coil, to change, along with the QFM surface shape. The free parameters are the surface shape Fourier coefficients $R_{\mathrm{b}, mn}, Z_{\mathrm{b}, mn}$ and the converter coil shape coefficients $X_n, Y_n, Z_n$ and coil current. The results are presented in figure~\ref{fig:SLNAM-to-SLAM}.


Adding the magnetic mirror coilset and axisymmetrizing changes the boundary and significantly alters the physics properties of the resulting fields compared to the initial OP equilibrium. To understand this effect, we must study the physics properties of equilibria corresponding to the new flux surface shapes $\mathcal{S}_1$ and $\mathcal{S}_2$. However, the racetrack equilibrium, or any boundary with an elongated axis shape, cannot currently be solved with~\texttt{DESC}, as parametrizing an elongated axis in cylindrical coordinates is suboptimal. We demonstrate this problem in~\ref{app:DESC-vs-GVEC}. To overcome this problem, we calculate the boundary data in different straight-field-line coordinate systems by tracing a single field line.



\subsection{Mapping flux surfaces by tracing a single field line with the $\texttt{FTGen}$ algorithm}
\label{sec:FTGen}

Adding the mirror section to an OP coilset and axisymmetrization changes the magnetic field and, consequently, the shape of the outermost nested flux surface. To calculate an ideal-MHD equilibrium for the resulting hybrid configuration pre- and post-axisymmetrization, we require a parametrization of the new boundary.
Although the QFM optimization described in the previous section provides a geometric surface with small normal magnetic-field error, the QFM surface
itself does not provide the rotational transform $\iota$, poloidal current $G$ or a straight-field-line parametrization. We instead obtain all three directly from the magnetic field
generated by the final coilset using a novel field-line-tracing surface-generation algorithm, which we refer to as~\texttt{FTGen} (Flux Tube Generator).

The central idea of $\texttt{FTGen}$ is that a sufficiently long trajectory of a single field line on an irrational flux surface samples the entire toroidal
surface\footnote{The idea of mapping a two-dimensional quantity on a surface from a field-aligned, one-dimensional quantity was motivated by the ballooning transform \cite{ConnorHastieTaylorProcRoySoc, DewarGlasserballooning}.}. Rather than treating the trace as a one-dimensional curve, $\texttt{FTGen}$ uses the rotational transform to assign successive toroidal transits of the field
line different field-line labels. Consider a field line traced through an unwrapped, uniformly-spaced geometric toroidal angle $\hat{\zeta} \in [0, 2\pi n_{\rm{transit}})$, which we decompose as
\begin{equation}
    \hat{\zeta} = 2\pi k + \zeta, \qquad k \in \{0, 1, \dots, n_{\rm{transit}}-1\},
\end{equation}
where $k$ is the toroidal transit number and $\zeta \in [0,2\pi)$ is the uniformly-spaced geometric toroidal angle within a single transit. The magnetic axis is first determined from a field-line trace by finding the field-line starting position for which $R(\zeta = 2 \pi) = R(\zeta=0), Z(\zeta = 2\pi) = Z(\zeta=0)$ using a root-finding algorithm. Then we map the flux surface of interest starting on the field line $\alpha = 0$, and using the magnetic axis location define the geometric poloidal angle
\begin{equation}
\hat{\theta}(\hat{\zeta}, \alpha = 0) =
    \mathrm{arctan}\left(
    \frac{Z(\hat{\zeta})-Z_{\rm{axis}}{(\hat{\zeta})}}
    {R(\hat{\zeta})-R_{\rm{axis}}(\hat{\zeta})}\right),
    \quad \hat{\theta} \in [0, 2\pi \iota n_{\rm{transit}}),
\end{equation}
where the rotational transform $\iota$ is unknown. However, knowing $\hat{\theta}$ and $\hat{\zeta}$ allows us to calculate the rotational transform on the flux surface,
\begin{equation}
    \iota \simeq
    \frac{\Delta \hat{\theta}}{\Delta \hat{\zeta}}.
\end{equation}
In practice, we trace the field line for $n_{\rm{transit}} = 96$--$256$, with the optimal $n_{\rm{transit}}$ depending on $\iota$ and hence on how rapidly the
field line samples the surface. Since the initial field line is chosen to have $\alpha_0=0$, its $k^{\rm{th}}$ toroidal transit corresponds to $\alpha_k = 2\pi\iota k\pmod{2\pi}$ and the $\alpha_k$, in general, will not be in any particular order. Therefore, after sorting $\alpha$, we obtain all the quantities of interest $R, Z, X, Y, \bi{B}$ on the ordered grid $(\alpha, \zeta)$ with a uniformly-spaced $\zeta$ but a non-uniformly spaced $\alpha$. 
It is also useful to additionally construct a Boozer-coordinate representation of the surface. We define the Boozer toroidal angle as
\begin{equation}
    \zeta_B
    =
    \zeta+\nu ,
\end{equation}
where $\nu$ is the Boozer toroidal-angle shift. For the vacuum magnetic field
generated by the coilset, $\nu$ is obtained directly along the initial
field-line trace by integrating the streamline equation $d\hat{\zeta}_{B}/(\bi{B}\cdot \bi{\nabla}\hat{\zeta}_{B}) = d\hat{\zeta}/(\bi{B}\cdot \bi{\nabla}\hat{\zeta})$ and using the fact that for a vacuum equilibrium in Boozer coordinates $\bi{B} \cdot \bi{\nabla} \hat{\zeta}_B = B^2/G$. After simplification, the streamline equation becomes an ODE 
\begin{equation}
    \frac{d\nu}{d\hat{\zeta}} = \frac{B^2}{G\,(\bi{B}\cdot\bi{\nabla}\hat{\zeta})}-1 ,
    \label{eqn:streamline-equation}
\end{equation}
that can be integrated on a flux surface given a scalar $G$, the poloidal current in Boozer coordinates. Every quantity except $G$ is known on the right side of~\eref{eqn:streamline-equation}. $\texttt{FTGen}$ determines $G$ by requiring the integrated shift $\nu$ to be bounded and centered over the trace,
\begin{equation}
    \max_{\hat{\zeta}}\nu(\hat{\zeta};G) + \min_{\hat{\zeta}}\nu(\hat{\zeta};G) = 0.
    \label{eqn:nu-condition}
\end{equation}
By solving~\eref{eqn:nu-condition} as a root-finding problem, we can calculate $G$ and subsequently $\nu(\hat{\zeta})$. In general, any physical quantity will be non-periodic in $\zeta$ for a given $\alpha$ in the $(\alpha, \zeta)$ coordinate system. Hence, we fit 2D splines to the data in $(\alpha, \zeta)$ coordinates using~\texttt{scipy.interpolate.RectBivariateSpline}.
Effectively, we have sampled every quantity on a single field line and mapped it as a two-dimensional surface quantity,
\begin{equation}
    F(\hat{\zeta}_{k,j})
    \longrightarrow
    F(\alpha_k,\zeta_j), \quad F \in \left\{R,Z,X,Y,\bi{B},\nu \right\}.
\end{equation}

\begin{figure}[h]
    \centering
    \begin{subfigure}[b]{0.55\textwidth}
    \centering
        \includegraphics[width=\textwidth, trim={1mm 2mm 0mm 1mm}, clip]{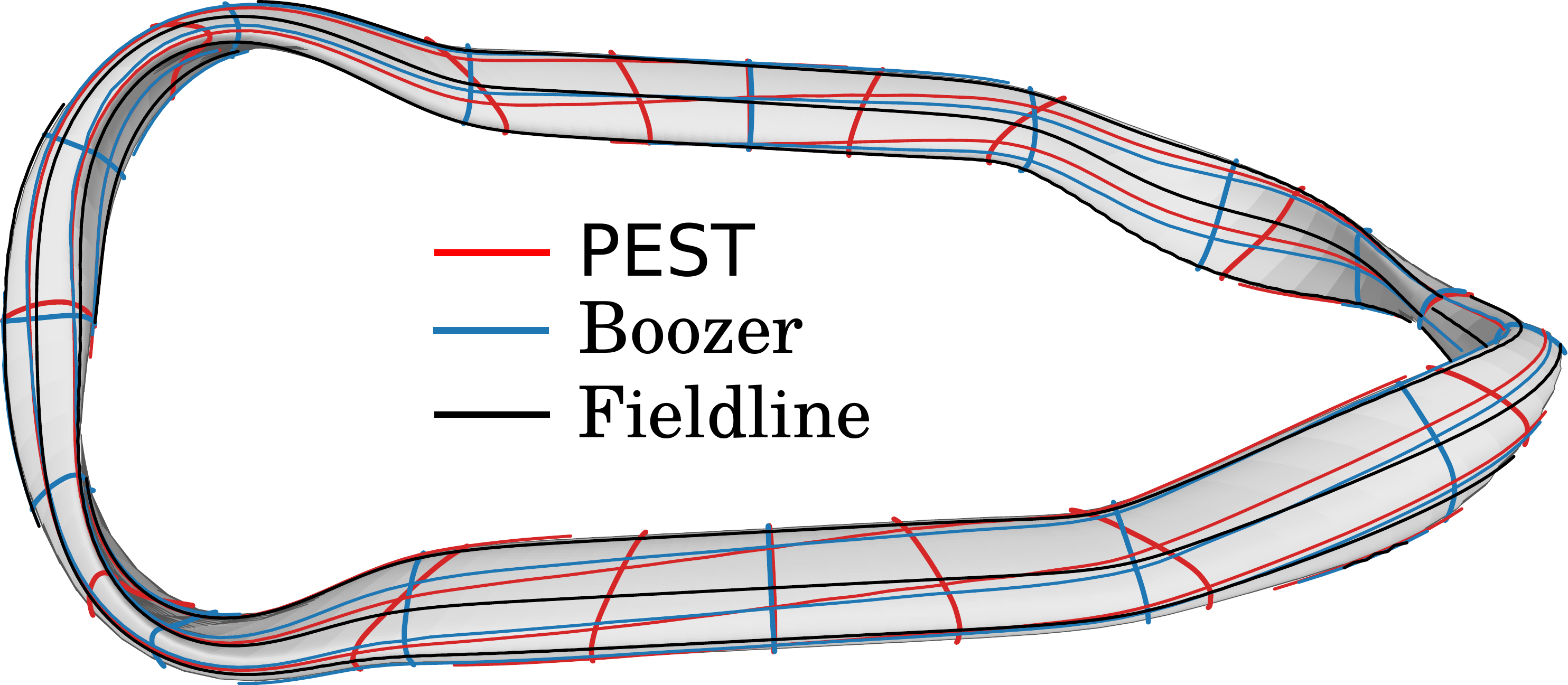}\\[6mm]
        \caption{Grids on the boundary}
        \label{subfig:alpha-phi}
    \end{subfigure}
    \begin{subfigure}[b]{0.38\textwidth}
        \centering
        \includegraphics[width=\textwidth, trim={0mm 2mm 0mm 0mm}, clip]{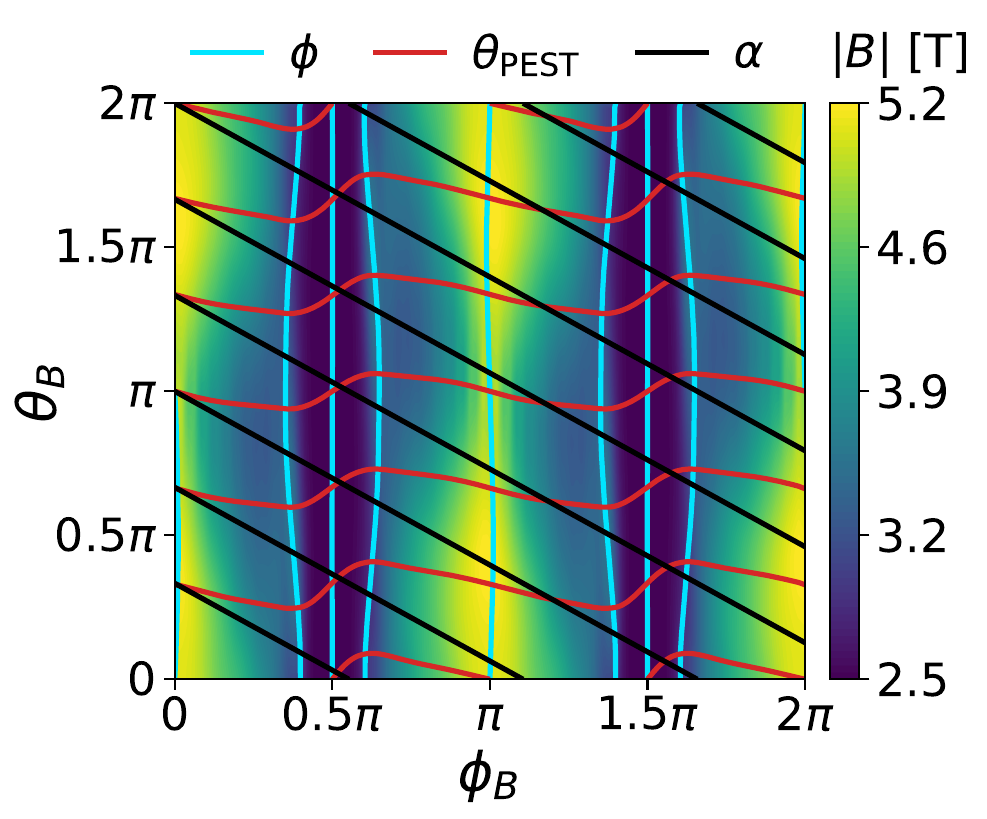}
        \caption{$|\bi{B}(\theta_{B}, \zeta_{B})|$}
        \label{subfig:theta_B-phi_B}
    \end{subfigure}
    \caption{In figure~\ref{subfig:alpha-phi}, we present a pre-SLAM surface mapped using the~\texttt{FTGen} algorithm showing grids corresponding to different coordinate systems. Contours of constant $\zeta_{B}$ result in a cross-section that is approximately normal to the mirror region whereas contours of constant cylindrical angle lead to an oblique cross-section. In figure~\ref{subfig:theta_B-phi_B}, we plot the $|\bi{B}|$ on the surface in Boozer coordinates and contours of constant PEST coordinates  and the field-line label $\alpha$.}
\label{fig:FTGen} 
\end{figure}

The first step is to use the definition $\alpha = \theta_{\rm{PEST}} - \iota \zeta$ and perform an affine transformation from $(\alpha, \zeta)$ to $(\theta_{\rm{PEST}}, \zeta)$. This allows us to calculate all the quantities on an ordered uniform PEST grid $(\theta_{\rm{PEST}}, \zeta)$ by evaluating $F(\theta_{\rm{PEST}}-\iota \zeta, \zeta)$. Note that all the quantities are periodic in PEST coordinates. The second step is to evaluate all the quantities $F$ on a uniform, ordered grid in Boozer coordinates $(\theta_{B}, \zeta_{B})$. To do that, for each prescribed point $(\theta^{*}_{B}, \zeta^{*}_{B})$ we use $\nu(\theta_{\rm{PEST}}, \zeta)$ to solve the root-finding problem 
\numparts
\begin{eqnarray}
     \zeta^{*}_B &= \zeta + \nu(\theta_{\rm PEST},\zeta),\\
     \theta^{*}_B &= \theta_{\rm PEST} + \iota\nu(\theta_{\rm PEST},\zeta),
\end{eqnarray}
\endnumparts
giving the PEST coordinates on which we can then evaluate $F(\theta_{\rm{PEST}}-\iota \zeta, \zeta)$, finally obtaining all the physical quantities on the prescribed Boozer grid points $(\theta_B, \zeta_B)$. We demonstrate the application of the $\texttt{FTGen}$ algorithm in figure~\ref{fig:FTGen}.

Thus, the $\texttt{FTGen}$ algorithm gives us all the physical quantities $F$ in three different coordinate systems $(\alpha, \zeta)$, $(\theta_{\rm{PEST}}, \zeta)$, and $(\theta_{B}, \zeta_{B})$. This construction is useful for stellarator-mirror hybrid configurations because it allows us to parametrize a flux surface in multiple coordinate systems, given a coilset. In particular, the method is different from the QFM surface approach, where one parametrizes a surface in a double Fourier basis and solves for the coefficients along with the scalars $\iota$ and $G$, first developed by Giuliani \etal~\cite{giuliani2022boozersurfaces}. For SLAMs, our method is more modular and robust than the one developed by Giuliani \etal. 

The resulting surface data from $\texttt{FTGen}$ are provided to the $\texttt{GVEC}$ fixed-boundary equilibrium solver to calculate an ideal MHD equilibrium. The process is described in detail in the following section.

\section{Calculating the ideal MHD hybrid equilibria using GVEC}
\label{sec:GVEC}
In this section, we will use the flux surface data in Boozer coordinates to parametrize and solve the fixed-boundary equilibrium using the Galerkin Variational Equilibrium Code (\texttt{GVEC})~\cite{Hindenlang2026_GVEC, Hindenlang2025_GFrame}. 
Unlike~\texttt{DESC}, which uses global spectral coefficients representing the boundary shape with the toroidal angle $\zeta$ fixed as the cylindrical angle $\phi$, \texttt{GVEC} defines the position vector on a flux surface as $\bi{x}=h\bigl(X^{1}(\rho,\hat{\theta},\hat{\zeta}), X^{2}(\rho,\hat{\theta},\hat{\zeta}), \hat{\zeta}\bigr)$, where $\hat{\theta}$ and $\hat{\zeta}$ can in principle be any poloidal and toroidal angle. Only $X^{1}$ and $X^{2}$ are solved for while the map $h$ is chosen by the user and held fixed~\cite{Hindenlang2025_GFrame}. With the generalized coordinate frame (G-frame),
\begin{equation}
  h=\bi{X}_{0}(\hat{\zeta})+X^{1}\bi{N}_{1}(\hat{\zeta})
   +X^{2}\bi{N}_{2}(\hat{\zeta}),
\end{equation}
the cross-sections, i.e. the $\zeta=\mathrm{const}$ planes, follow a closed guiding curve $\bi{X}_{0}$. Here $\bi{N}_{1}$ and
$\bi{N}_{2}$ are differentiable vectors that span each cross-sectional plane~\cite{Hindenlang2025_GFrame}.

For SLAMs, the difference in geometric representation is decisive.  Since the~\texttt{DESC} formulation is limited to the cylindrical toroidal angle $\phi$, a straight mirror $\phi=\mathrm{const}$ plane cuts the axis obliquely everywhere except at the section midpoint. A circular mirror cross-section appears as an ellipse whose elongation grows toward the mirror ends, and the extra shaping must be carried by additional Fourier harmonics, making convergence to an equilibrium state difficult (see~\ref{app:DESC-vs-GVEC}). This becomes worse as the ratio $L/D$ increases. In an axis-following frame, the same cross-section is always a circle for a SLAM. 
In~\texttt{GVEC}, we can use a G-frame to parametrize the boundary, rather than the cylindrical toroidal angle $\phi$. This frame is constructed by averaging the Boozer toroidal angle, following~\cite{Hindenlang2025_GFrame}.

\begin{figure}[h]
    \vspace*{-1mm}
    \centering
    \begin{subfigure}[b]{0.45\textwidth}
    \centering
        \includegraphics[
        width=0.82\textwidth, trim={2mm 38mm 8mm 5mm}, clip]{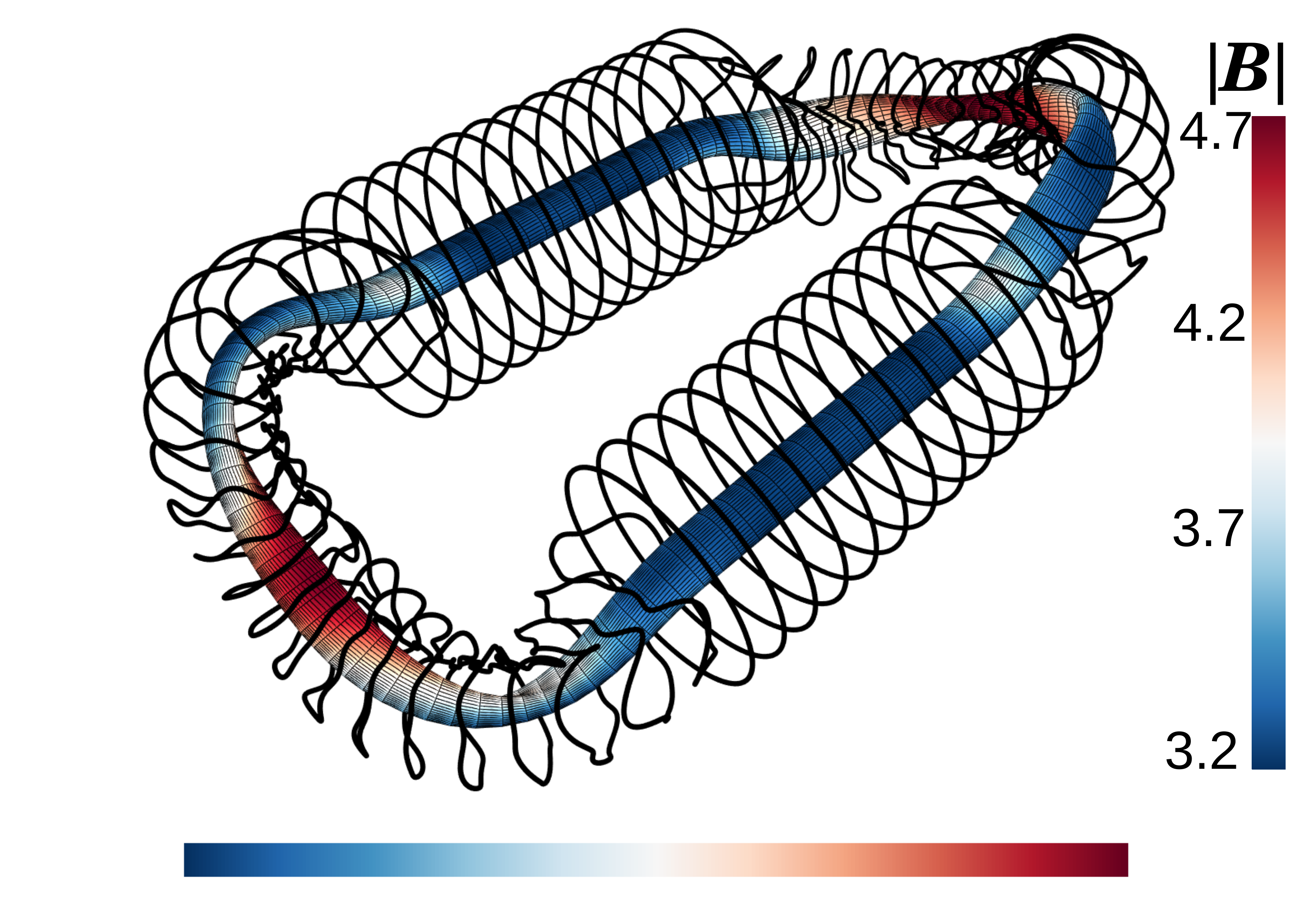}
    \end{subfigure}
    \begin{subfigure}[b]{0.45\textwidth}
        \centering
        \raisebox{2.0mm}{
        \includegraphics[width=0.8\textwidth, trim={2mm 30mm 8mm 5mm}, clip]{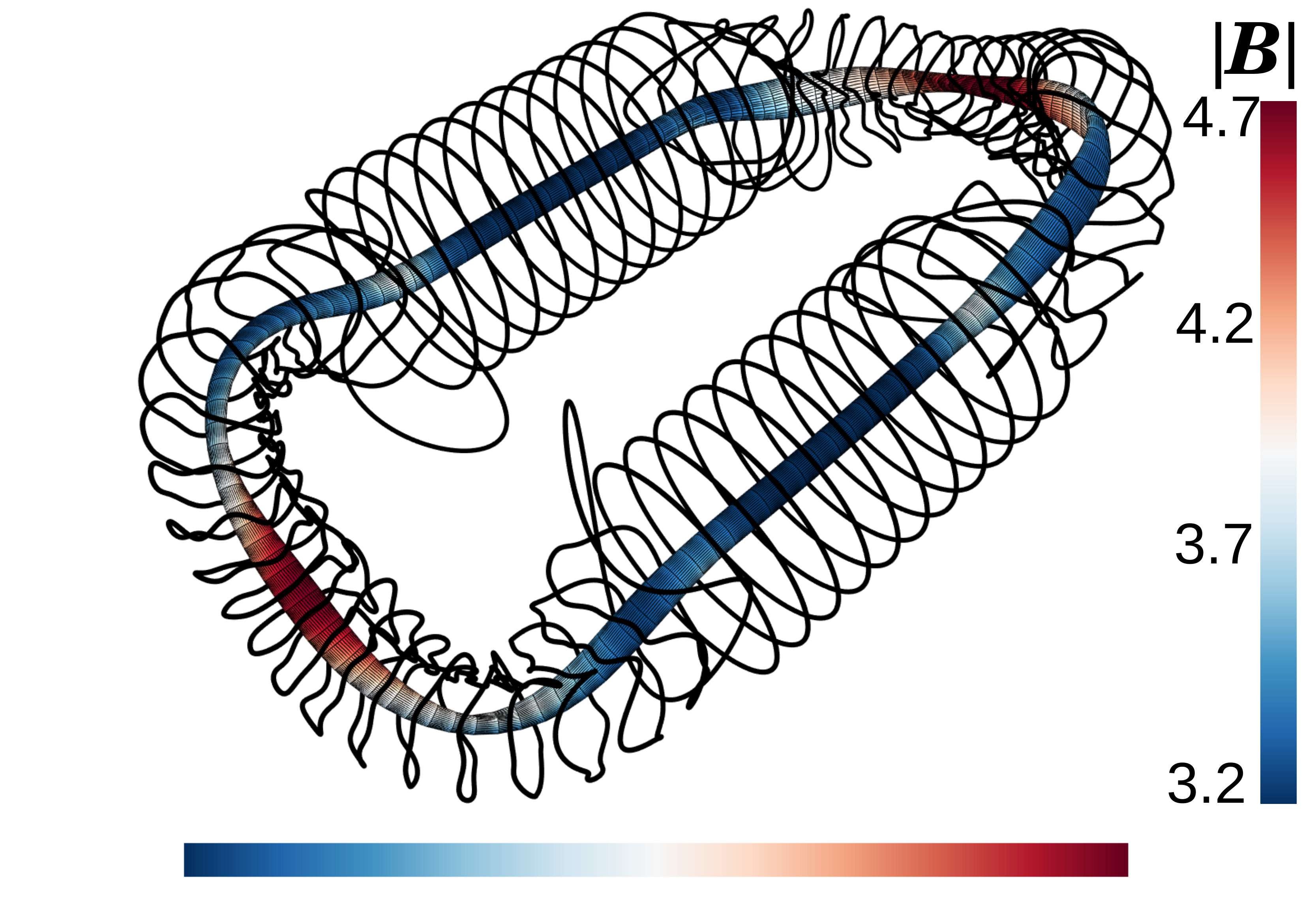}}
    \end{subfigure}\\
    
    \begin{subfigure}[b]{0.47\textwidth}
    \centering
        \raisebox{5.0mm}{
        \includegraphics[width=\textwidth, trim={2mm 2mm 2mm 2mm}, clip]{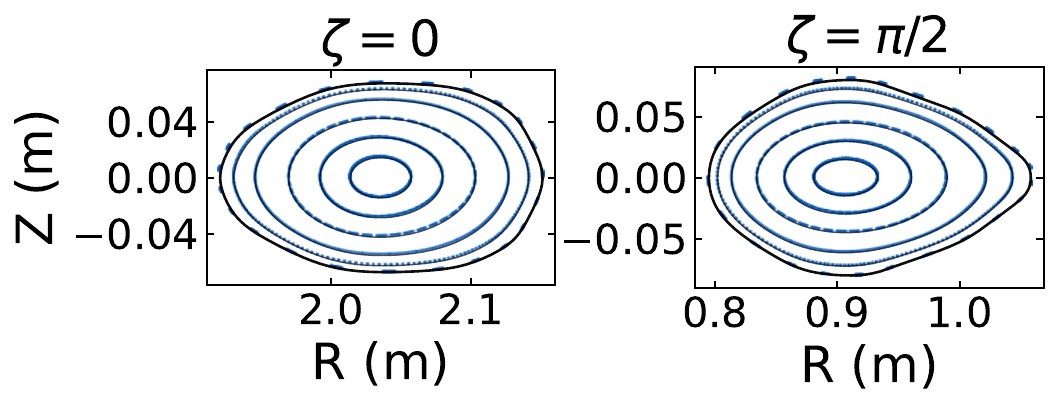}}
        \caption{pre-SLAM}
        \label{subfig:SLNAM}
    \end{subfigure}
    \quad 
    \begin{subfigure}[b]{0.45\textwidth}
        \centering
        \raisebox{-30.0mm}{
        \includegraphics[width=\textwidth, trim={2mm 2mm 2mm 2mm}, clip]{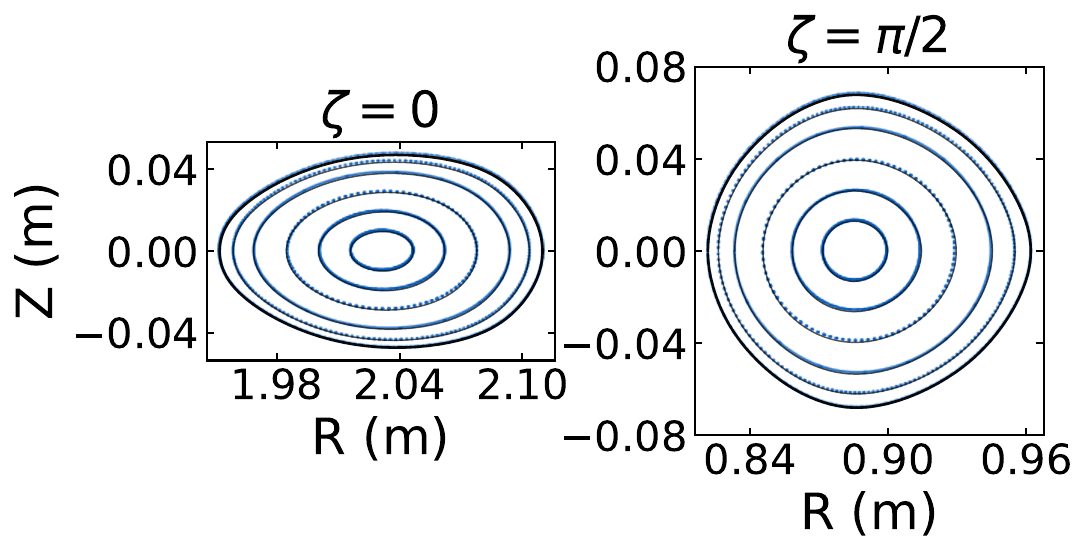}}
        \caption{SLAM}
        \label{subfig:SLAM}
    \end{subfigure}
    \caption{Comparison of the fixed-boundary~\texttt{GVEC} solution with the Poincar\'e trace from the coilset, demonstrating an excellent match between the~\texttt{GVEC} surfaces and the Poincar\'e data. Note that, even though the surfaces are nested at larger radii, the SLAM equilibrium was truncated to a smaller radius because the outer surfaces become non-circular in the mirror region, making the SLAM equilibrium smaller.}
\label{fig:GVEC-3D}
\end{figure}

After parametrizing the boundary using a G-frame, we run~\texttt{GVEC} to calculate the pre-SLAM and SLAM equilibria and present the results in figure~\ref{fig:GVEC-3D}.
We observe that the pre-SLAM and SLAM solutions match well with the field line traced from the coilset. This validates the ability of $\texttt{FTGen}$ to generate smooth high-resolution data that can be read by any fixed-boundary solver. The lowest field occurs in the mirror region and the highest occurs in the stellarator region, but the reduction is non-monotonic, which departs from the behavior of a poloidally omnigenous field. The non-monotonic change in the magnetic field strength is partly due to the ripple introduced by the converter coil, as shown in figure~\ref{subfig:NFP3-hybrid}. 

\begin{figure}[h]
    \centering
    \begin{subfigure}[b]{0.32\textwidth}
    \centering
        \includegraphics[width=\textwidth, trim={1mm 2mm 0mm 0mm}, clip]{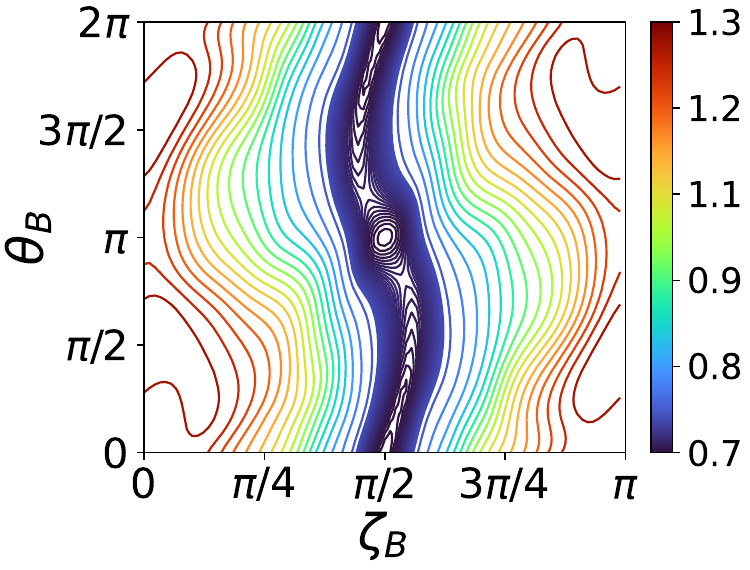}
        \caption{OP ($\rho=1$)}
        \label{subfig:OP}
    \end{subfigure}
    \! 
    \centering
    \begin{subfigure}[b]{0.32\textwidth}
    \centering
        \includegraphics[width=\textwidth, trim={0mm 2mm 0mm 0mm}, clip]{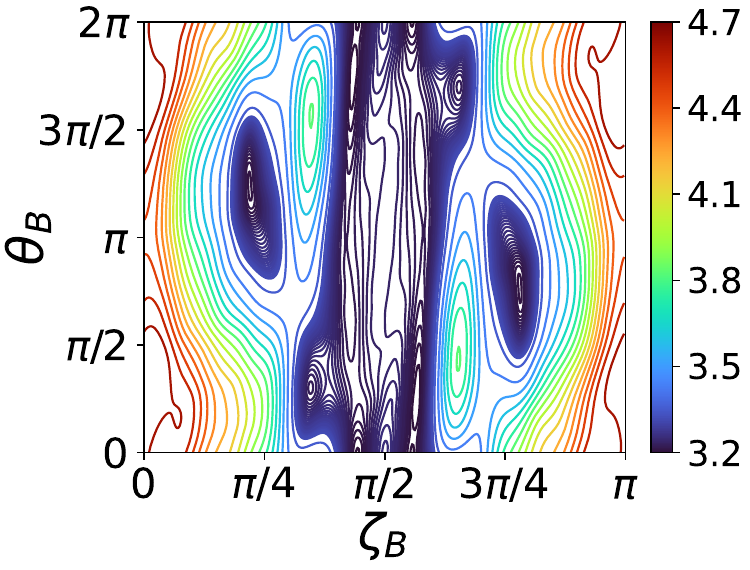}
        \caption{pre-SLAM ($\rho=1$)}
        \label{subfig:pre-SLAM}
    \end{subfigure}
    \!
    \begin{subfigure}[b]{0.32\textwidth}
        \centering
        \includegraphics[width=\textwidth, trim={0mm 2mm 0mm 0mm}, clip]{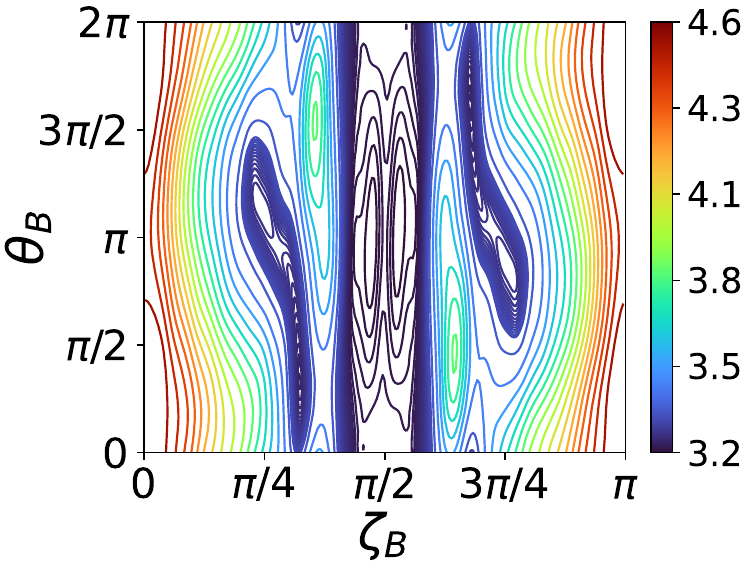}
        \caption{SLAM ($\rho=1$)}
        \label{subfig:SLAM-Boozer}
    \end{subfigure}
    \caption{Magnetic field strength on the boundary in Boozer coordinates for the precursor OP, pre-SLAM, and SLAM equilibria on their respective boundaries. The mirror region lies in the low-field region. Using an axisymmetrized mirror brings the low-field region closer to a OP configuration by closing or straightening more puddles and making more $|\bi{B}|$ contours closed in $\theta_{\rm{B}}$ compared to the pre-SLAM case.}
    \vspace*{-4mm}
\label{fig:Boozer_contours} 
\end{figure}

To further understand the differences, we plot the magnetic field strength on the boundary for the three equilibria in Boozer coordinates in figure~\ref{fig:Boozer_contours}.
We find that axisymmetrizing the mirror closes more $|\bi{B}|$ contours poloidally, a necessary condition for omnigenity, making the mirror region closer to a QP configuration compared to the pre-SLAM case. Therefore, axisymmetrizing the mirror can be used as an objective to approach an omnigenous SLAM configuration.

This concludes our procedure for designing a SLAM. We started from an OP coilset, created a stellarator-mirror coilset, traced the boundary from the hybrid coilset using the $\texttt{FTGen}$ algorithm and fed it to~\texttt{GVEC} to obtain a fixed-boundary equilibrium. This is the reverse of the conventional order of designing a stellarator. In the next section, we will analyze the physics properties of the OP, pre-SLAM, and the corresponding SLAM equilibrium.

\subsection{Neoclassical transport, MHD stability, kinetic turbulence}
\label{sec:physics}
\begin{figure}[h]
    \centering
    \begin{subfigure}[b]{0.32\textwidth}
    \centering
        \includegraphics[width=\textwidth, trim={2mm 2mm 0mm 0mm}, clip]{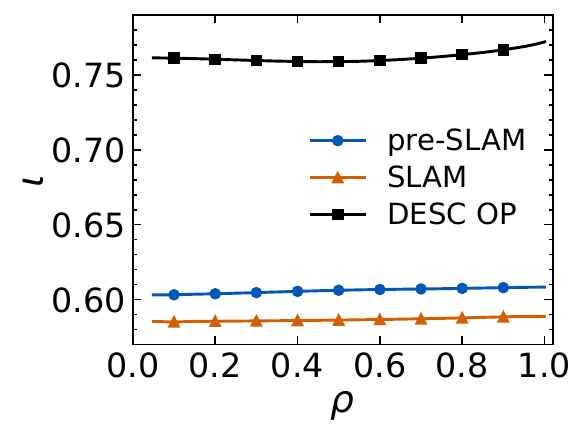}
        \label{subfig:iota}
    \end{subfigure}
    \! 
    \centering
    \begin{subfigure}[b]{0.32\textwidth}
    \centering
        \includegraphics[width=\textwidth, trim={2mm 2mm 0mm 0mm}, clip]{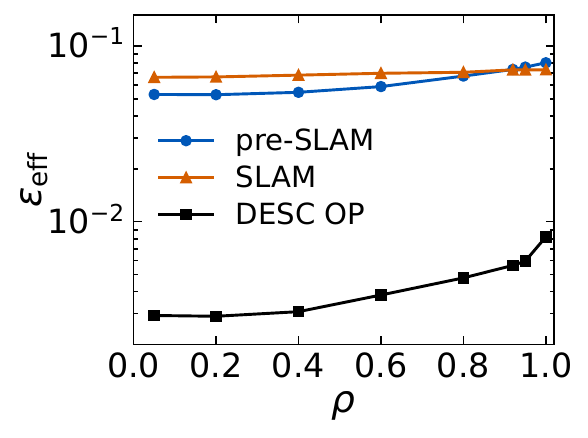}
        \label{subfig:Ripple}
    \end{subfigure}
    \!
    \begin{subfigure}[b]{0.32\textwidth}
        \centering
        \includegraphics[width=\textwidth, trim={2mm 2mm 0mm 0mm}, clip]{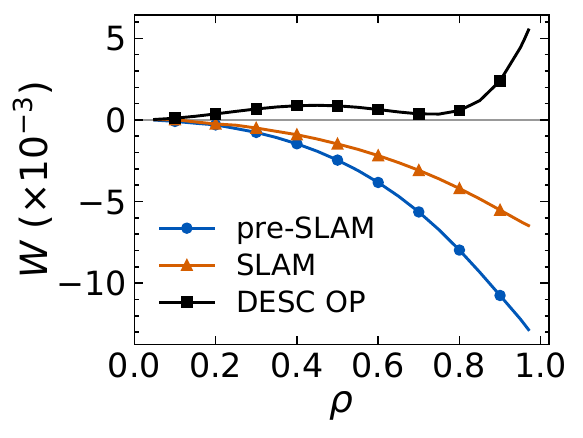}
        \label{subfig:Well}
    \end{subfigure}\\[-4mm]
    \caption{Comparison of the rotational transform, effective ripple, and the magnetic well between the original OP, pre-SLAM, and SLAM equilibria. Both the ripple and magnetic well are worse after converting the OP coilset into a stellarator-mirror hybrid coilset. The SLAM ripple is worse in the core but better in the edge, and its magnetic well is better than that of the corresponding pre-SLAM.}
\label{fig:Ripple-Well} 
\end{figure}

To understand the physics properties of the SLAMs, we compare the precursor OP, pre-SLAM, and SLAM designs by calculating various performance metrics. In this work, we will focus on three properties: the neoclassical transport proxy $\epsilon_{\rm{eff}}$ from Nemov \etal~\cite{nemov1999evaluation}, implemented in~\texttt{DESC} using the algorithm developed by Unalmis \etal~\cite{unalmis2026spectrally}, MHD stability described by the magnetic well, and the local gyrokinetic ITG heat flux using the~\texttt{GX} code~\cite{Mandell2024GX}. We first compare the effective ripple, which measures the neoclassical confinement in the $1/\nu$ regime, and the magnetic well $W$, which is a measure of stability against interchange modes, in figure~\ref{fig:Ripple-Well}.

We find that our prescription for converting an OP coilset into a stellarator-mirror hybrid does not preserve any of the physical properties for which the original OP was optimized. The SLAM equilibrium has the worst ripple of all three, which is most likely due to the converter coil creating a large ripple in the converter region, as shown in figure~\ref{subfig:modB-on-axis}. Axisymmetry of the mirror comes at the cost of increased ripple.

\begin{figure}[h]
    \centering
    \begin{subfigure}[b]{0.35\textwidth}
    \centering
        \includegraphics[width=\textwidth, trim={0mm 2mm 0mm 0mm}, clip]{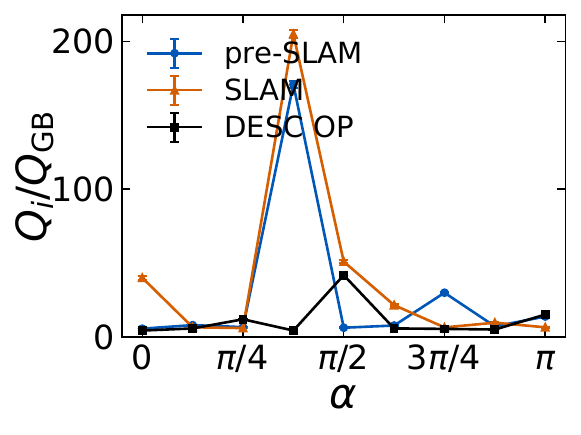}
        \label{subfig:GX-alpha}
    \end{subfigure}
    \qquad 
    \begin{subfigure}[b]{0.35\textwidth}
        \centering
        \includegraphics[width=\textwidth, trim={0mm 2mm 0mm 0mm}, clip]{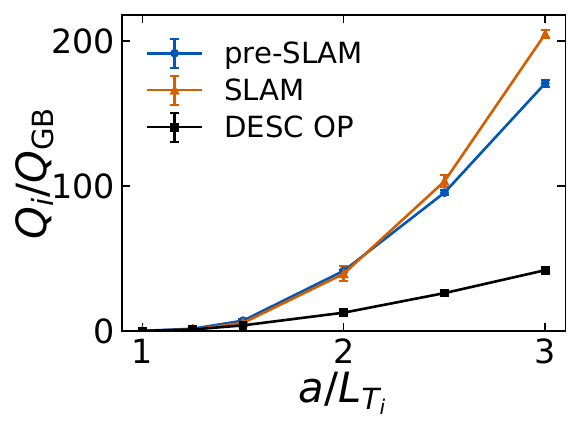}
        \label{subfig:GX-aLT}
    \end{subfigure}\\[-6mm]
    \caption{Nonlinear ITG heat flux $Q$(in gyro-Bohm units $Q_{\rm{GB}}$) from~\texttt{GX} for the original OP, pre-SLAM, and SLAM equilibria at $\rho = 0.8$. On the left, $Q$ as a function of the field-line label $\alpha$ at a normalized ion-temperature gradient $a/L_T = 3$. On the right, we plot $Q$ as a function of $a/L_T$ on the field line with the largest heat flux.}
\label{fig:GX} 
\end{figure}

Next, we perform local gyrokinetic turbulence calculations for all three equilibria using the~\texttt{GX} code. 
We solve the single-species nonlinear ion gyrokinetic equation with adiabatic electrons on a single flux surface $\rho = 0.8$ at five different field lines $\alpha \in \{0, \pi/5, 2\pi/5, 3\pi/5, 4\pi/5\}$ for a field-line length $z \in [-5\pi, 5\pi]$. 
This gives us the heat flux due to ion-temperature-gradient (ITG) turbulence. Next, we perform these calculations for multiple values of the ion-temperature gradient. We present our findings in figure~\ref{fig:GX}.

\begin{figure}[h]
    \centering
    \begin{subfigure}[b]{0.49\textwidth}
    \centering
        \includegraphics[width=\textwidth, trim={0mm 2mm 0mm 0mm}, clip]{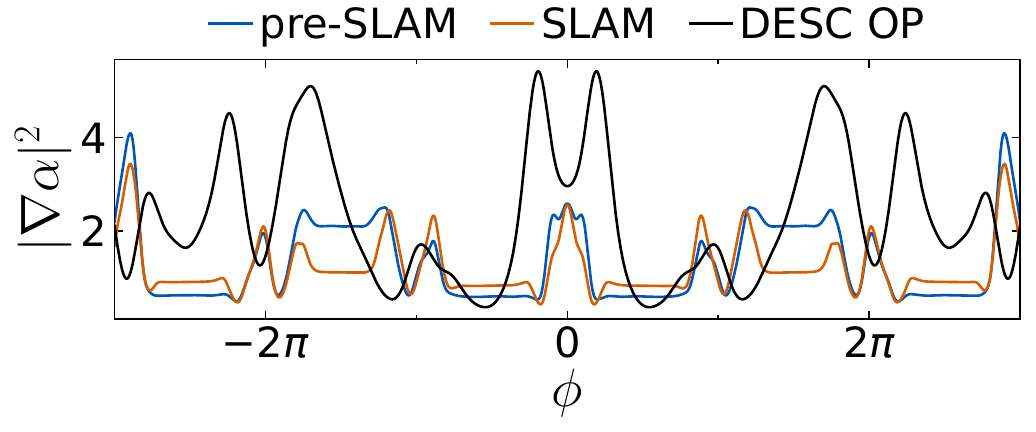}
        \label{subfig:grad-alpha}
    \end{subfigure}
    \! 
    \begin{subfigure}[b]{0.49\textwidth}
        \centering
        \includegraphics[width=\textwidth, trim={0mm 2mm 0mm 0mm}, clip]{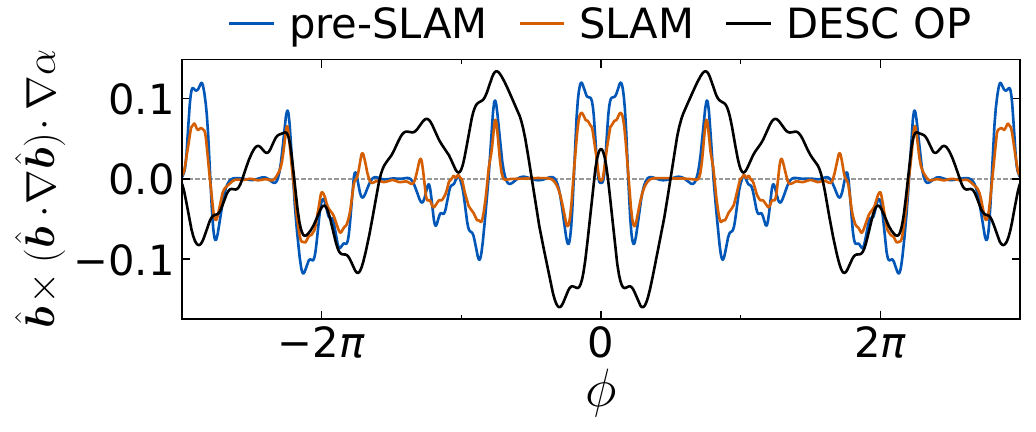}
        \label{subfig:cvdrift}
    \end{subfigure}\\[-6mm]
    \caption{Geometric coefficients along the flux tube on the field line $\alpha = 0$ at $\rho = 0.8$ for the OP, pre-SLAM, and SLAM equilibria. In the mirror region of both hybrids, $|\bi{\nabla}\alpha|^2$ is flat and the curvature drift nearly vanishes, a characteristic property of stellarator--mirror hybrid equilibria.}
\label{fig:geometry} 
\end{figure}

As none of the equilibria were optimized against ITG turbulence, we find that increasing the temperature gradient significantly increases the heat flux for all three equilibria. Moreover, the heat flux from the stellarator-mirror hybrids is considerably larger than the precursor OP with the heat flux peaking in the converter region between the stellarator and mirror. To understand the difference between the OP and the hybrids, figure~\ref{fig:geometry} shows two geometric coefficients of the~\texttt{GX} flux tube, $|\bi{\nabla}\alpha|^2$, which sets the perpendicular wavenumber and hence the finite-Larmor-radius (FLR) stabilization, and the curvature-drift coefficient $\hat{\bi{b}}\times(\hat{\bi{b}}\cdot\bi{\nabla}\hat{\bi{b}})\cdot\bi{\nabla}\alpha$.
Two features distinguish the hybrids from the OP. First, $|\bi{\nabla}\alpha|$ is nearly constant in $\phi$ throughout the mirror region. The axisymmetric mirror contributes almost nothing to the rotational transform, so the local pitch of the field line, and with it $|\bi{\nabla}\alpha|$, barely changes there, as is also visible in figure~\ref{subfig:alpha-phi}. Second, the mirror axis is straight, so the mirror region has no toroidal curvature and the curvature drift nearly vanishes. These geometric coefficients, together with the mirror length that sets the parallel extent of this curvature-free region, can therefore be tuned to keep the net curvature favorable and the FLR stabilization strong, similar to the critical-gradient optimizations work by Roberg-Clark et al.~\cite{RobergClark2021_LinearCriticalGradient, roberg2022coarse, RobergClark2026_QICriticalGradient}. To summarize, we find that a high-performance SLAM will require optimization against turbulence to reach an acceptable level of transport. 

\section{Summary and conclusions}
\label{sec:conclusions}

We have presented a computational design study of Stellarators Linking Axisymmetric Mirrors (SLAMs), a stellarator--mirror hybrid in which a poloidally omnigenous (OP) stellarator supplies the rotational transform that closes the loss cone of long, axisymmetric magnetic mirrors. The motivation is the volumetric cost-scaling argument of section~\ref{sec:cost-benefit}. At equal plasma volume, the cost ratio $C_r$ of a conventional stellarator to a SLAM increases from $1/\hat{\gamma}$ toward $\hat{\alpha}$ as the mirror-to-stellarator volume ratio $V_r = M_r L_r$ grows, so that a hybrid is worth pursuing because axisymmetric mirror sections are cheaper per unit plasma volume than three-dimensional stellarator sections. The present work addresses the baseline SLAM-A, with the mirror inserted at the natural field minimum of the OP precursor and a mirror ratio of order unity.

Rather than following the conventional stage-one or stage-two order, we designed the device starting from the coils. Taking an $\NFP = 2$ OP coilset from the~\texttt{DESC} omnigenity database, we separated it at the low-field region, where the axis curvature vanishes, and filled the gap with planar circular mirror coils. Field-line tracing shows that nested vacuum flux surfaces survive this extrusion for mirror lengths at least up to $L/D = 2$ and for mirror-coil currents between $0.8$ and $2.0$ times that of the adjacent stellarator coil, with the low-field cross-section of the OP precursor qualitatively preserved. To address the result that the mirror cross-section of the resulting pre-SLAM is not circular, we re-optimized a Quadratic Flux Minimizing (QFM) surface together with a single converter coil per half field period, using the new shaping objectives (\ref{app:objectives}), to make the mirror section circular and collimated along a straight axis while leaving the stellarator section and its $|\bi{B}|$ unchanged. To obtain a straight-field-line boundary parametrization with the rotational transform $\iota$ and poloidal current $G$ that a QFM surface does not supply, we introduced the $\texttt{FTGen}$ algorithm, which maps a single traced field line onto ordered $(\alpha,\zeta)$, PEST, and Boozer grids. Since a racetrack-shaped axis is poorly represented in the cylindrical toroidal angle (\ref{app:DESC-vs-GVEC}), the Boozer-parametrized boundary was passed to the axis-following G-frame of~\texttt{GVEC}, whose fixed-boundary pre-SLAM and SLAM equilibria reproduce the coil Poincar\'e traces closely. The same procedure was applied to an $\NFP = 3$ precursor in~\ref{app:NFP3}.

We find that converting the OP coilset into a hybrid degrades every property for which the precursor was optimized: the effective ripple $\epsilon_{\rm{eff}}^{3/2}$ and the magnetic well are both worse for the pre-SLAM and SLAM than for the OP, the SLAM has the largest core ripple of the three, which can be explained by the observation that the converter coil introduces a large $|\bi{B}|$ ripple in the converter region, and the on-axis field is no longer monotonic between the stellarator maximum and the mirror minimum, which departs from the necessary condition needed for poloidal omnigenity. Axisymmetrizing the mirror does, however, improve the magnetic well and the edge ripple relative to the pre-SLAM. Nonlinear ITG heat fluxes from~\texttt{GX} rise steeply with the ion-temperature gradient in all three cases, but the increase in the flux is a lot more significant in the hybrid configurations. 

The principal conclusion is that the coil-first construction presented here can be used as a robust initialization procedure for SLAMs rather than as an optimized design. Nested surfaces, a circular mirror, and a converged equilibrium are obtained, but the confinement and stability properties of the OP precursor are not inherited. The workflow developed here naturally extends in several directions. Finite plasma pressure can be included immediately, since~\texttt{GVEC} takes a pressure profile as input, allowing the $\beta$ scaling $\beta_m = M_r^2 \beta_s$ and finite-$\beta$ MHD stability~\cite{gaur_2024_13887566, gaur2023adjoint, gaur2026agni} to be tested.  A high-performance SLAM will require direct stage-one optimization of the full hybrid boundary with the mirror region included in the objectives, which in turn requires a generalized toroidal-angle boundary representation of the kind available in~\texttt{GVEC} or a corresponding extension of~\texttt{DESC}. Design and optimization of SLAMs is left as future work.

\textbf{Data availability}
The data, analysis, and post-processing files used in this work are publicly available at \url{https://doi.org/10.5281/zenodo.23158822}~\cite{gaur_2026_23158822}.

\textbf{Acknowledgements}
The authors gratefully acknowledge fruitful discussions with B. Jang, R. Conlin, C. Zhu, and E. Balkovich. This research used resources of the Perlmutter supercomputer located at the
National Energy Research Scientific Computing Center, a DOE Office of Science User Facility supported by the Office of Science of the U.S. Department of Energy under contract No.
DE-AC02-05CH11231 using NERSC award FES-ERCAP0035536.
The contributions of R. Babin and F. Hindenlang have been carried out within the framework of the EUROfusion Consortium, funded by the European Union via the Euratom Research and Training Programme (Grant Agreement No 101052200 — EUROfusion). Views and opinions expressed are however those of the authors only and do not necessarily reflect those of the European Union or the European Commission. Neither the European Union nor the European Commission can be held responsible for them.

\appendix
\section{Stellarator mirror hybrids with $n_{\mathrm{FP}} > 2$}
\label{app:NFP3}
\begin{figure}[h]
    \vspace*{-5mm}
    \centering
    \begin{subfigure}[b]{0.39\textwidth}
    \centering
        \includegraphics[width=\textwidth, trim={75mm 30mm 0mm 0mm}, clip]{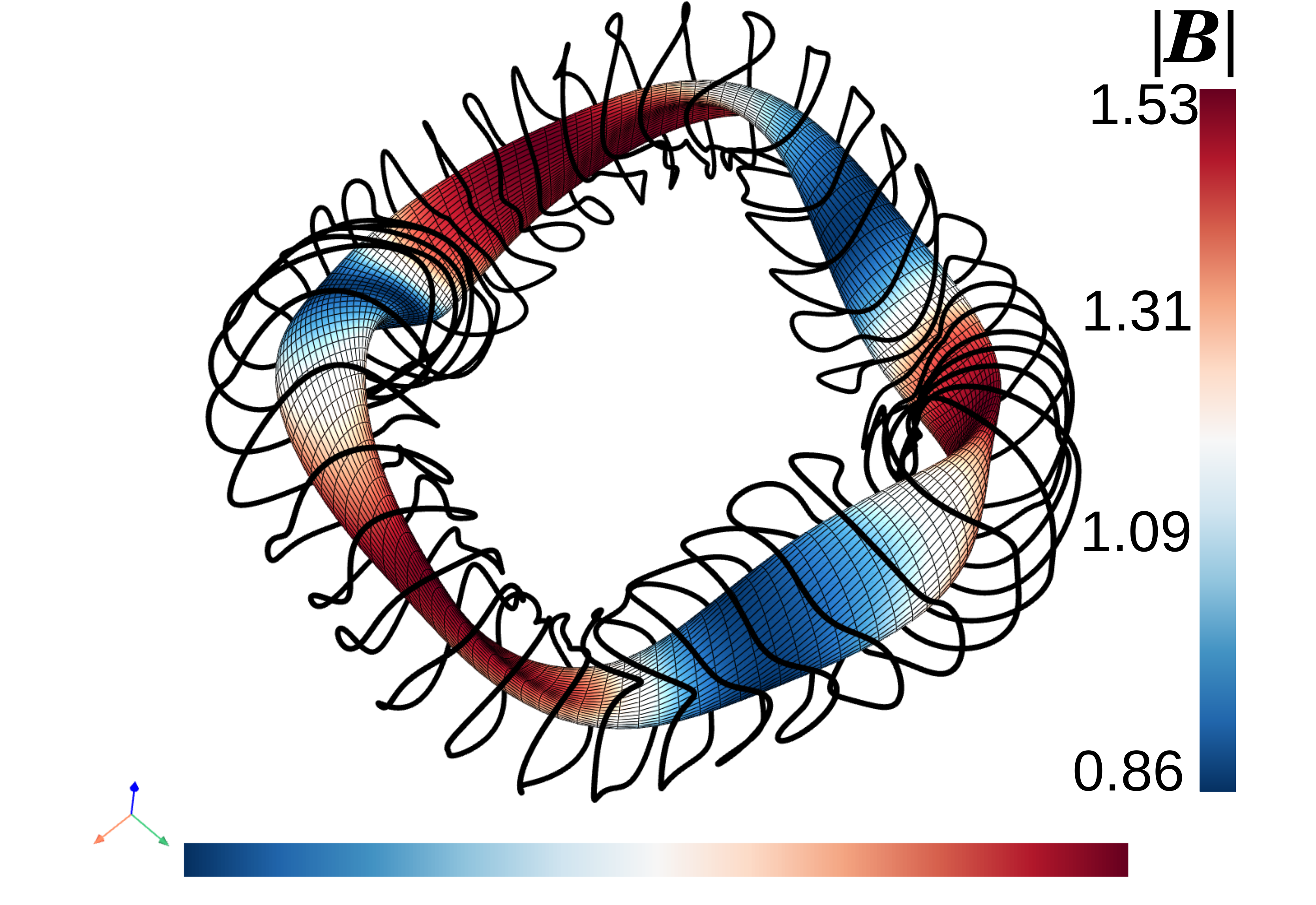}
        \caption{Precursor OP}
        \label{subfig:NFP3-OP}
    \end{subfigure}
    \begin{subfigure}[b]{0.49\textwidth}
        \centering
        \includegraphics[width=\textwidth, trim={40mm 58mm 0mm 0mm}, clip]{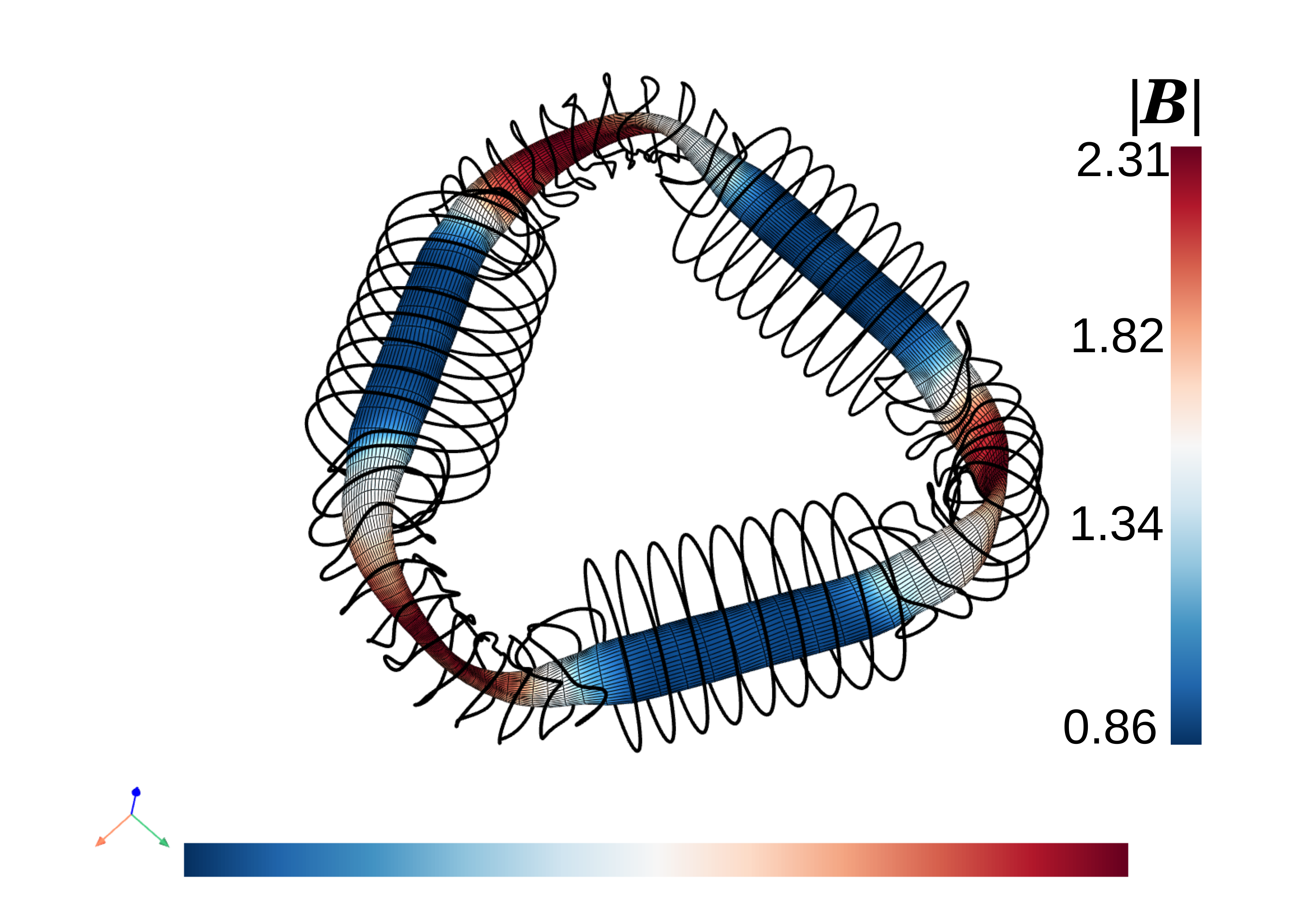}
        \caption{Hybrid}
        \label{subfig:NFP3-hybrid}
    \end{subfigure}
    \caption{An $\NFP=3$ precursor OP in figure~\ref{subfig:NFP3-OP} calculated in~\texttt{DESC} and a corresponding stellarator-mirror hybrid in figure~\ref{subfig:NFP3-hybrid} calculated in~\texttt{GVEC}. The choice of mirror coil current causes the mirror ratio $B_{\max}/B_{\min}$ to differ from the precursor OP equilibrium.}
\label{fig:NFP3} 
\end{figure}

In this appendix, we present an $\NFP=3$~\texttt{GVEC} equilibrium created from an $\NFP=3$ OP equilibrium using the method described in this paper. The coilset and magnetic field strength $| \bi{B} |$ on the boundary are provided in figure~\ref{fig:NFP3}.
We used the procedure explained in this paper to create the stellarator-mirror hybrid coilset, followed by the~\texttt{FTGen} algorithm to obtain flux surface data which was then fed to~\texttt{GVEC}. This approach is flexible and robust for any $\NFP > 1$ vacuum stellarator equilibrium.

\section{The cylindrical angle $\phi$ is sub-optimal for racetrack shapes}
\label{app:DESC-vs-GVEC}
\begin{figure}[h]
    \vspace*{-2mm}
    \centering
    \begin{subfigure}[b]{0.49\textwidth}
    \centering
        \includegraphics[width=\textwidth, trim={2mm 2mm 0mm 0mm}, clip]{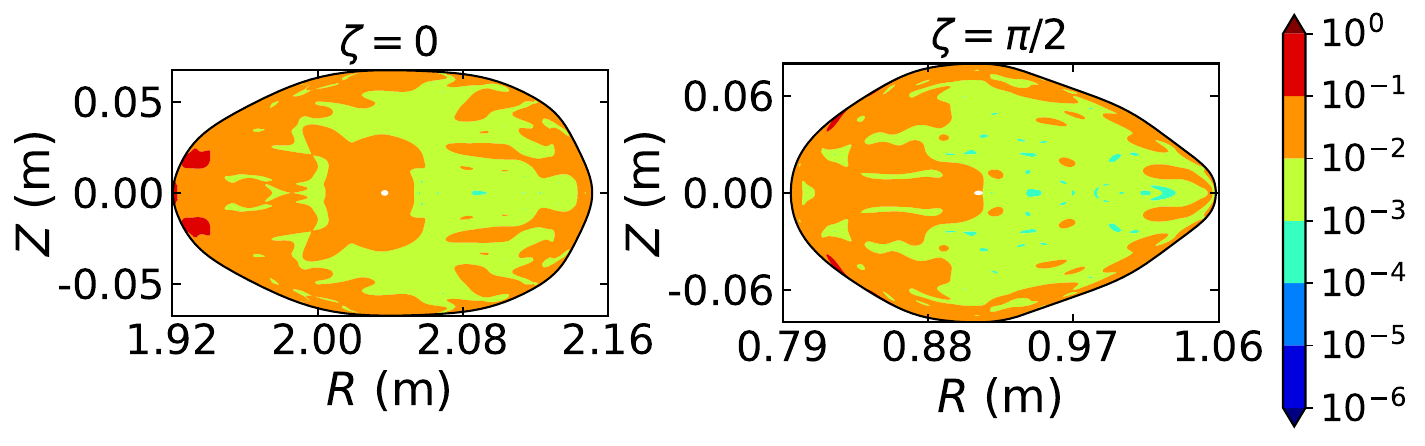}
        \caption{DESC}
        \label{subfig:normF_DESC}
    \end{subfigure}
    \begin{subfigure}[b]{0.49\textwidth}
        \centering
        \includegraphics[width=\textwidth, trim={2mm 2mm 0mm 0mm}, clip]{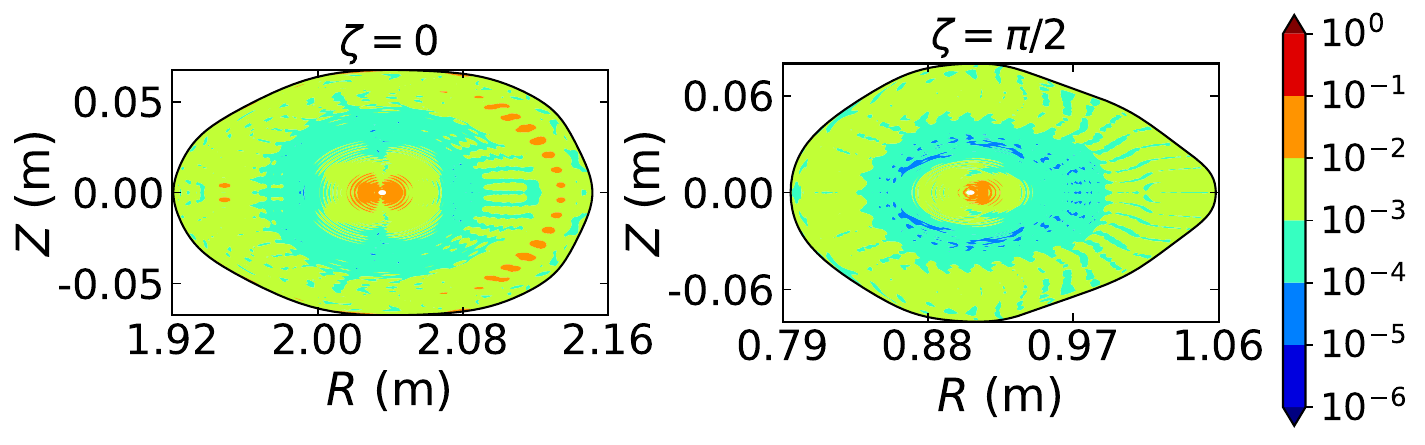}
        \caption{GVEC}
        \label{subfig:normF_GVEC}
    \end{subfigure}
    \caption{Comparison of the normalized force balance error at two different cross-sections obtained by solving the equilibrium for the pre-SLAM boundary corresponding to $\mathcal{S}_1$ in figure~\ref{subfig:SLNAM-QFM}. The boundary data was calculated by tracing the field line using the~\texttt{FTGen} algorithm and fed to both codes. The~\texttt{DESC} case solved the equilibrium problem in native~\texttt{DESC} coordinates whereas~\texttt{GVEC} solved the problem in generalized coordinates.}
\label{fig:normF-SLAM-SLNAM} 
\end{figure}

In this appendix, we solve the pre-SLAM equilibrium using~\texttt{DESC} and~\texttt{GVEC}. Both codes take the flux surface data generated using the~\texttt{FTGen} algorithm, but~\texttt{DESC} parametrizes the boundary using the cylindrical angle whereas~\texttt{GVEC} parametrizes the boundary in a generalized toroidal angle constructed from Boozer coordinates. We present a comparison of the normalized force balance error $|F|_{\rm{normalized}} = |\bi{j} \times \bi{B} - \bi{\nabla}p|/\langle (B^2/2\mu_0)\rangle_{\rm{vol}}$, where $\langle \rangle_{\rm{vol}}$ is a volume average operator, in figure~\ref{fig:normF-SLAM-SLNAM}. 

The only difference between the input boundary data used by~\texttt{DESC} and~\texttt{GVEC} is the coordinate system used to parametrize the boundary. This demonstrates that using generalized curve-following coordinates allows~\texttt{GVEC} to achieve a better equilibrium, as~\texttt{DESC} is currently limited to using the cylindrical toroidal angle, which yields an equilibrium with poor force balance. A generalized toroidal angle feature in~\texttt{DESC} would be advantageous for SLAMs.

\section{SLAM shaping objectives}
\label{app:objectives}
In this appendix, we describe the new objectives designed to axisymmetrize a stellarator-mirror hybrid. Before explaining the individual objectives in detail, we clarify some symbols and conventions. The pre-SLAM QFM surface $\mathcal{S}_1$ and SLAM QFM surface $\mathcal{S}_2$ are parametrized in a double Fourier basis in $(\theta,\zeta)$, with $\zeta=\phi$ the cylindrical toroidal angle, so that $X=R\cos\zeta$ and $Y=R\sin\zeta$. Each objective is evaluated on a grid of collocation nodes $(\theta_k,\zeta_l)$, $k=1,\dots,M$, $l=1,\dots,L$, and returns a residual vector $\mathbf{f}$. The optimizer minimizes $\frac{1}{2}\sum_{\rm{obj}} w_{\rm{obj}}^2\|(\mathbf{f}_{\rm{obj}}-\mathbf{f}^\star_{\rm{obj}})/\mathcal{N}_{\rm{obj}}\|_2^2$, where $w$ is the weight, $\mathbf{f}^\star$ the target and $\mathcal{N}$ a normalization constant.
 
\subsection{Circular mirror cross-section, $f_{\mathrm{SLAM-elong}}$}
The purpose of this objective is to enforce circularity in a non-circular mirror section. As a plane of constant toroidal angle $\zeta$ in~\texttt{DESC} would not be perpendicular to the mirror axis, we devise an objective that provides us with the true cross-section, by calculating the cross-section perpendicular to the mirror center line. In general, the mirror center line is not horizontal as it lies in the $X$--$Z$ plane with the form $Z = -s X$ ($s = -dZ/dX$).
The planes perpendicular to the center line are described as $X-sZ=X_{0,j}$, $j=1,\dots,J$. We define the coordinate normal to these planes, $u$, and the in-plane coordinate, $v$,
\begin{equation}
  u=\frac{X-sZ}{\sqrt{1+s^2}},\qquad v=\frac{sX+Z}{\sqrt{1+s^2}},\qquad u_{0,j}=\frac{X_{0,j}}{\sqrt{1+s^2}},
\end{equation}
so that plane $j$ is $u=u_{0,j}$ and $(Y,v)$ are Cartesian coordinates within it.
 
A curve of constant $\theta_k$ on the surface runs toroidally and crosses plane $j$ once. The grid nodes on that curve that lie within a slab of half-width $\sim\sigma$ about that plane are averaged with Gaussian weights normalized along $\zeta$,
\begin{equation}
\fl  \omega_{kl}^{(j)}=\frac{\exp\!\bigl[-(u_{kl}-u_{0,j})^2/2\sigma^2\bigr]}
                         {\sum_{l'=1}^{L}\exp\!\bigl[-(u_{kl'}-u_{0,j})^2/2\sigma^2\bigr]},\qquad
  y_k=\sum_{l=1}^{L}\omega_{kl}^{(j)}Y_{kl},\qquad z_k=\sum_{l=1}^{L}\omega_{kl}^{(j)}v_{kl},
\end{equation}
where $\sigma=b\,(\max_{k,l}u_{kl}-\min_{k,l}u_{kl})$ and $b$ is a user-set bandwidth. Because the weights sum to one along each $\theta_k$ curve, the band of nodes inside the slab corresponds to a single point $(y_k,z_k)$ per $\theta_k$. The $M$ points $(y_k,z_k)$ form one ordered closed polygon per plane. The Gaussian allows us to work with a polygon with a fixed number of sides, which makes the objective differentiable. Using cyclic indexing ($(k \bmod M)+1$) and defining $c_k=y_kz_{k+1}-y_{k+1}z_k$, the shoelace formula gives the polygon area and area centroid,
\begin{equation}
\fl  \mathcal{A}=\frac12\sum_{k=1}^{M}c_k,\qquad
  y_c=\frac{1}{6\mathcal{A}}\sum_{k=1}^{M}(y_k+y_{k+1})c_k,\quad
  z_c=\frac{1}{6\mathcal{A}}\sum_{k=1}^{M}(z_k+z_{k+1})c_k .
\end{equation}
The distance of each vertex from the centroid, its mean, and the normalized radius are
\begin{equation}
 \fl r_k=\sqrt{(y_k-y_c)^2+(z_k-z_c)^2},\qquad \bar r=\frac1M\sum_{k=1}^{M}r_k,\qquad
  s_k= \frac{r_k}{\bar r}.
\end{equation}
To calculate the elongation, we calculate the ratio of the maximum to minimum radius as a smooth function using the log-mean-exp function,
\begin{equation}
\fl  s_{\max}=\frac{1}{\hat{s}}\ln\Bigl(\frac1M\sum_{k=1}^{M}e^{\hat{s} s_k}\Bigr),\qquad
  s_{\min}=-\frac{1}{\hat{s}}\ln\Bigl(\frac1M\sum_{k=1}^{M}e^{-\hat{s} s_k}\Bigr),\qquad
  E_j=s_{\max}/s_{\min},
\end{equation}
where $\hat{s}$ is a sharpness parameter such that for $\hat{s} \rightarrow \infty$, $s_{\max} \rightarrow \max_k s_k$ and $s_{\min} \rightarrow \min_k s_k$. The objective is the largest elongation over all planes,
\begin{equation}
  f_{\mathrm{SLAM-elong}}=\max_{j}E_j,\qquad f^\star_{\mathrm{SLAM-elong}}=1 .
\end{equation}
Since $\frac{1}{M}\sum_k s_k=1$ by construction, Jensen's inequality gives $s_{\max}\ge1$ and $s_{\min}\le1$, with equality if and only if every $s_k=1$. Hence $E_j=1$ if and only if $r_k=\bar r$ for all $k$,
\textit{i.e.}, the cross-section is a circle about its centroid.

\subsection{Straight mirror center line, $f_{\mathrm{SLAM-center}}$}
This objective ensures that the center of the mirror region is a straight line and does not assume a finite curvature. At each coordinate $X_{0,j}$ along the straight section we want to check if it is centered on the mirror coil axis. To make the objective differentiable, we use a Gaussian weight, exactly as in $f_{\mathrm{SLAM-elong}}$,
\begin{equation}
  K_{kl}^{(j)}=\exp\!\Bigl[-\frac{(X_{kl}-X_{0,j})^2}{2\sigma_X^2}\Bigr]\mathcal{J}_{kl},\qquad
  \sigma_X=b\,(\max X-\min X),
\end{equation}
where $\mathcal{J}_{kl}=|\partial_\theta\mathbf{x}\times\partial_\zeta\mathbf{x}|$ is the area Jacobian, included so that it takes into account the inhomogeneity of the grid points. The area-weighted centroid of that soft slice is
\begin{equation}
  \bar Y_j=\frac{\sum_{kl}K^{(j)}_{kl}Y_{kl}}{\sum_{kl}K^{(j)}_{kl}},\qquad
  \bar Z_j=\frac{\sum_{kl}K^{(j)}_{kl}Z_{kl}}{\sum_{kl}K^{(j)}_{kl}},
\end{equation}
\textit{i.e.},\ $(\bar Y_j,\bar Z_j)$ is our estimate of where the plasma cross-section is centered at the $j$-th section. The mirror coils have centers that trace out a straight line $Z=Z_c(X)$ at fixed $Y = Y_c$. We want the plasma centroid to sit on that same line at every section, so the residual is simply the offset from it,
\begin{equation}
\fl  \mathbf{f}_{\mathrm{SLAM-center}} = \Bigl[\bigl(\bar Y_j-Y_{0,j}\bigr)_{j=1}^{J},\; \bigl(\bar Z_j-Z_{0,j}\bigr)_{j=1}^{J}\Bigr],
  \qquad \mathbf{f}^\star=\mathbf{0},\qquad \mathcal{N}=R_0 ,
\end{equation}
with $Y_{0,j}=Y_c$ and $Z_{0,j}=Z_c(X_{0,j})$. At $\mathbf{f}=\mathbf{0}$, the plasma cross-section is centered on the coil axis at every sampled section, \textit{i.e.}, the mirror is collimated with the line joining the coil centers.

\subsection{Matching the stellarator region, $f_{\rm{surf-match}}$}
The purpose of this objective is to match the SLAM stellarator section with the pre-SLAM stellarator section. Let $\mathcal{S}_1$, with coordinates $(R^{(1)},Z^{(1)})$, be the pre-SLAM QFM surface, held fixed, and let $C_S$ be the collocation nodes lying in the stellarator region. We want $\mathcal{S}_2$, the surface being optimized, to coincide with $\mathcal{S}_1$ there. The residual is the signed difference at each node,
\begin{equation}
\fl  \mathbf{f}_{\rm{surf-match}}=
  \Bigl[\bigl(R(\theta_k,\zeta_l)-R^{(1)}(\theta_k,\zeta_l)\bigr),\;
        \bigl(Z(\theta_k,\zeta_l)-Z^{(1)}(\theta_k,\zeta_l)\bigr)\Bigr]_{(\theta_k,\zeta_l)\in C_S},
  \qquad \mathbf{f}^\star=\mathbf{0},
\end{equation}
not the distance $\sqrt{\Delta R^2+\Delta Z^2}$, since the distance has an undefined gradient if $\mathcal{S}_2$ starts as a copy of $\mathcal{S}_1$. The normalization is $\mathcal{N}=a^{(1)}$ (effective minor radius), computed from $\mathcal{S}_1$. At $\mathbf{f}=\mathbf{0}$,
$\mathcal{S}_2$ coincides with $\mathcal{S}_1$ node for node.

\newpage
\section*{References}
\bibliographystyle{ieeetr}
\bibliography{main}

\end{document}